\documentclass[%
reprint,
superscriptaddress,
 amsmath,
 amssymb,
 aps,
]{revtex4-1}

\usepackage{appendix}
\usepackage{graphicx}
\usepackage{dcolumn}
\usepackage{bm}
\usepackage{gensymb}
\usepackage{lipsum}
\usepackage{amsfonts}
\usepackage{amsmath}
\usepackage{amssymb}
\usepackage{epstopdf}
\usepackage{algorithmic}
\usepackage{color}
\usepackage{amsopn}

\newcommand{\jt}[1]{{\color{black} #1}}
\newcommand{\jw}[1]{{\color{black} #1}}

\begin{document}

\title{Trajectory stratification of stochastic dynamics}

\author{Aaron R. Dinner}
\email{dinner@uchicago.edu}
\affiliation{James Franck Institute, The University of Chicago, Chicago, Illinois 60637, USA}
\affiliation{Department of Chemistry, The University of Chicago, Chicago, Illinois 60637, USA}

\author{Jonathan C. Mattingly}
\affiliation{Departments of Mathematics and Statistical Science, Duke University, Durham, North Carolina 27708, USA}

\author{Jeremy O. B. Tempkin}
\affiliation{James Franck Institute, The University of Chicago, Chicago, Illinois 60637, USA}
\affiliation{Department of Chemistry, The University of Chicago, Chicago, Illinois 60637, USA}

\author{Brian Van Koten}
\affiliation{Department of Statistics, The University of Chicago, Chicago, Illinois 60637, USA}

\author{Jonathan Weare}  
\email{weare@uchicago.edu}
\affiliation{James Franck Institute, The University of Chicago, Chicago, Illinois 60637, USA}
\affiliation{Department of Statistics, The University of Chicago, Chicago, Illinois 60637, USA}

\begin{abstract}
We present a general mathematical framework for trajectory stratification for
simulating rare events. Trajectory stratification involves decomposing
trajectories of the underlying process into fragments  limited to restricted
regions of state space (strata), computing averages over the distributions of
the trajectory fragments within the strata with minimal communication between
them, and combining those averages with appropriate weights to yield averages
with respect to the original underlying process. Our framework reveals the full
generality and flexibility of trajectory stratification, and it illuminates a
common mathematical structure shared by existing algorithms for sampling rare
events.  We demonstrate the power of the framework by defining strata in terms
of both points in time and path-dependent variables for efficiently estimating
averages that were not previously tractable.
\end{abstract}

\maketitle

\section{Introduction}
\label{sec:intro}

Computer simulation is a powerful tool for the study of physical processes. \jt{Specifically, stochastic simulation methods have broad applicability in modeling physical systems in a variety of fields including chemistry, physics,  climate science, engineering, and economics \cite{Asmussen2007, Gardiner2009}. In many practical applications, the statistical properties of the process of interest are approximated by averages over many independent realizations of trajectories of the process, or, in the case of ergodic properties, by averages taken over a single very long trajectory of the process.}
However, for many systems, the most interesting events occur
infrequently and are therefore very difficult to observe by direct numerical
integration of the equations governing the dynamics.  For example, in chemistry,
the conformational changes responsible for the function of many molecules and,
in climate science,  extreme events like severe droughts and violent hurricanes,
occur on timescales orders of magnitude longer than the timestep for numerical
integration. This basic observation has motivated the development of numerous
techniques aimed at enhancing the sampling of rare events of interest without
sacrificing statistical fidelity (see \cite{Frenkel2002} for an account within
the context of molecular simulation).

In this article, we depart from standard enhanced sampling approaches and
develop a general mathematical and computational framework for the estimation of
statistical averages involving rare trajectories of stochastic processes.  Our
approach can be viewed as a form of stratified sampling, long a cornerstone of
experimental design in statistics (e.g., \cite{Neyman1934}).  In stratified
sampling, a population is divided into subgroups (strata), averages within those
strata are computed separately, and then averages over the entire state space
are assembled as weighted sums of the strata averages.   Stratification
also has a long history in computer simulations of condensed-phase systems as
umbrella sampling (US) \cite{Torrie1977, Pangali1979, Chandler1987, Frenkel2002,
Lelievre2010}. The key idea behind any stratified sampling  strategy is that,
when the strata are chosen appropriately, their statistics can be obtained
accurately with relatively low effort and combined to estimate the average of
interest with (much) less overall effort than directly sampling the stochastic
process to the same statistical precision. Here we show that the  trajectories
of an arbitrary discrete-time Markov process (including many dynamics with
memory, so long as they can be written as a suitable mapping) can also be
stratified:  they can be decomposed into fragments restricted to regions of
trajectory space (strata), averages over the distributions of trajectory
fragments within the strata can be computed with limited communication between
them, and  those averages can be combined in a weighted fashion to yield a very
broad range of statistics that characterize the dynamics.

These basic features are at the core of  the existing nonequilibrium umbrella
sampling (NEUS) method \cite{Warmflash2007, Dickson2009-2, Vanden-Eijnden2009},
which forms the starting point for our development.  NEUS was originally
introduced to estimate stationary averages  with respect to a given, possibly
irreversible, stochastic process \cite{Warmflash2007}. Starting in \cite{Dickson2009-2,
Vanden-Eijnden2009} it was observed that the general NEUS approach was applicable to certain dynamic averages as well.  The basic NEUS approach has been been applied and further developed in subsequent articles \cite{Dickson2009-1,Dickson2010, Dickson2011,Xu2013} and in the Exact Milestoning scheme \cite{Bello-Rivas2015}, which was derived from the  Milestoning method \cite{Faradjian2004}  but is very similar in structure to NEUS.
At its most basic level, NEUS relies on duplication of
states in rarely visited regions of space and subsequent forward evolution of
the duplicated states.  In this way it is similar to a long list of so-called
``trajectory splitting'' techniques \cite{Glasserman:1998:splitting,  Huber1996, HarasztiTownsend:1999:DPR, vanErp2003, Allen2005, JohansenDelMoral:2006:SMCrare,  CerouGuyader:2007:AMS, Guttenberg2012, HairerWeare:2014:TDMC}
 that are also
able to compute averages of dynamic quantities. Like NEUS, splitting techniques
also often involve a decomposition of state space into regions.  Unlike NEUS
however, in most splitting techniques bias is removed through the use of a
separate weight factor for each individual sample (rather than for an entire
region), and the computational effort expended in each region is not controlled
directly.   What makes the NEUS method unique among splitting techniques is that
it is also a trajectory stratification strategy.

Our goal in this article is to provide a clear and general mathematical
framework for trajectory stratification that builds upon the NEUS method. In the process we clearly delineate the range of statistics that can be estimated by NEUS, including more general quantities  than  previously computed. Our analysis of the underlying mathematical structure of US \cite{Thiede2016,Dinner2016} has already
facilitated the derivation of a central limit theorem for US and a detailed
understanding of its error properties. Here, our framework reveals unanticipated
connections between the equilibrium and nonequilibrium US methods and places the
nonequilibrium algorithm within the well-studied family of stochastic
approximation methods \cite{Kushner2003}. The analysis leads to a practical
scheme that departs dramatically from currently available alternatives. We
demonstrate the use of trajectory stratification to compute a hitting time
distribution  as well as to compute the expectation of a path-dependent
functional that gives the relative normalization constants for two arbitrary,
user-specified \jt{unnormalized}  probability densities. 

\section{A Unified Framework}
\label{sec:outline}

In this section we present a framework that reveals the unified structure underlying umbrella sampling in both the  equilibrium and nonequilibrium case. In Section \ref{sec:eus}, we review the equilibrium approach \cite{Thiede2016, Dinner2016} to introduce terminology and the central eigenproblem in a context where the analogies to traditional umbrella sampling descriptions \cite{Torrie1977, Pangali1979, Chandler1987, Frenkel2002, Lelievre2010} are readily apparent. In Section \ref{sec:gen}, we present the nonequilibrium version of the algorithm and show how this interpretation results in a flexible scheme for computing dynamic averages.  As for  its equilibrium counterpart, an eigenproblem lies at the core of the nonequilibrium method.  This eigenproblem however, involves a matrix that depends on the desired eigenvector, introducing the need for a self-consistent iteration. In Section \ref{sec:fp}, we give a precise description of the fixed-point problem solved by this iteration and show that the algorithm is an example of a stochastic approximation strategy \cite{Kushner2003}. In  Section \ref{sec:erg} we specialize our development to the context of steady-state averages that motivated the original development of NEUS \cite{Warmflash2007}.

\subsection{Averages with Respect to a Specified Density}
\label{sec:eus}

Our presentation in this section follows \cite{Thiede2016}. We view umbrella sampling as a method to compute averages of the form
\begin{equation}\label{fave}
\int_{x \in \mathbb{R}^d} f(x) \pi(dx),
\end{equation}
where  $\pi$ is a known probability distribution and $d$ is the dimension of the underlying system (e.g., the total number of position coordinates for all atoms in a molecular system).  For example,  $\pi$ might be the canonical distribution, $\pi(dx)   \propto e^{-\beta V(x)}dx$ where $V$ is a potential energy function, $\beta$ is an inverse temperature, and $f$ might be 1 on some set $A$ and 0 elsewhere. In this case, $-\beta^{-1} \log \int f(x) \pi(dx)$ can be regarded as the free energy of the set $A.$ 

Note that in our notation $\pi$ is a probability measure on $\mathbb{R}^d$ and $dx$ is an infinitesimal volume element in $\mathbb{R}^d.$  If the distribution $\pi$ has a density function $p(x)$ then $\pi(A) = \int_{x\in A} p(x) dx$ and, in particular,  $\pi(dx) = p(x) dx.$  This more general notation  is  useful when we move to our description of the nonequilibrium umbrella sampling scheme. As an aid to the reader, we choose to introduce it in the simpler setting of this section.

Consistent with traditional implementations of US \cite{Pangali1979, Frenkel2002}, we divide the computation of the average in \eqref{fave} into a series of averages over local subsets of space.  More precisely, instead of directly computing averages with respect to $\pi,$ we  compute averages with respect to $n$ probability distributions, $\pi_j$, each of which concentrates probability in a restricted region of space (relative to $\pi$ itself) with the goal of eliminating or reducing barriers to efficient sampling associated with  $\pi$.  So that general averages with respect to $\pi$ can be assembled, the $\pi_j$ satisfy 
$
\pi = \sum_{j=1}^n z_j\, \pi_j
$
for a set of weights $z_j$ to be defined in a moment.  

To obtain the restricted distributions $\pi_j$ we can set
 \begin{equation}\label{piistat}
 \pi_j(dx) = \frac{\psi_j(x) \pi(dx)}{\int_{y \in \mathbb{R}^d} \psi_j(y) \pi(dy)},
 \end{equation}
 where the $\psi_j$ are non-negative user defined functions satisfying $ \sum_{j=1}^n \psi_j(x) = 1 $ for all $x$ (this last requirement is relaxed in \cite{Thiede2016}). For example, one might choose $ \psi_j(x) = \mathbf{1}_{A_j}(x) / \sum_{\ell =1}^n \mathbf{1}_{A_\ell}(x)$, where the $A_j$ are a collection of sets covering the space to be sampled, and, for any set $A_{j}$, the function $\mathbf{1}_{A_j}(x)$  is 1 if $x\in A_j$ and 0 otherwise.  

Note that $
\pi = \sum_{j=1}^n z_j\, \pi_j
$ is satisfied with 
 \begin{equation}\label{z}
 z_j = \int_{x \in \mathbb{R}^d}  \psi_j(x)\pi(dx)
 \end{equation}
 and that the average \eqref{fave} with respect to $\pi$
 can be reconstructed using the equation
 \begin{equation}
 \int_{x \in \mathbb{R}^d} f(x) \pi(dx) = \sum_{j=1}^n z_j\, \langle f\rangle_j  \label{avefexp},
 \end{equation}
with
 \begin{equation}\label{f}
 \langle f\rangle_j  = \int_{x \in \mathbb{R}^d}  f(x)  \pi_j(dx).
 \end{equation}
Here  $z_{j}$ is the statistical weight associated with each distribution $\pi_j$ and   $\langle f \rangle_{j}$ are the averages of the observable $f$ against $\pi_{j}$. From \eqref{avefexp} we see that if we can sample from the $\pi_j$ and compute the $z_j$ then we can compute averages with respect to $\pi.$  Since $\pi_j$ is known explicitly in this case, it can be sampled by standard means (e.g., Langevin dynamics or Metropolis Monte Carlo \cite{Frenkel2002}).

Our key observation underpinning the equilibrium umbrella sampling method is that the  $z_j$ themselves are functions of averages with respect to the local distributions $\pi_j$:
  \begin{equation}
z_j  = \sum_{i=1}^n z_i\, F_{ij}\label{eigeq1}\quad\text{and}\quad \sum_{j=1}^n z_j = 1,
 \end{equation}
 where
 \begin{equation}\label{F}
 F_{ij} = \int_{x \in \mathbb{R}^d} \psi_j(x) \pi_i(dx).
 \end{equation}
 The matrix $F$ is stochastic (i.e., has non-negative entries with rows that sum to 1) and \eqref{eigeq1}, which is written in matrix-vector form as
 \begin{equation}\label{eigeq2}
 z^\text{\tiny T} F = z^\text{\tiny T} \quad\text{and}\quad \sum_{j=1}^n z_j = 1,
 \end{equation}
 is an eigenproblem that can be solved easily for the vector $z$. 
 
 We now have a stratification scheme for computing the target average in \eqref{fave} by sampling from the distributions $\pi_{j}$. Operationally, the main steps are as follows.\\
\begin{enumerate}
\item Assemble $F$ defined in \eqref{F} (or the alternative in Appendix \ref{Fflux} below) and $\langle f\rangle_j $ defined in \eqref{f} by sampling from $\pi_j$ defined in \eqref{piistat}. \\ 
\item Solve the eigenvector equation \eqref{eigeq2} for $z$ defined in \eqref{z}.\\
\item  Compute the desired expectation via \eqref{avefexp}.\\
\end{enumerate}

The efficiency of this \emph{equilibrium} US scheme has been analyzed in detail elsewhere \cite{Thiede2016,Dinner2016}. Roughly, the benefit of US is due to the facts that averages with respect to the $\pi_j$ are often sufficient to solve for all desired quantities, and one can choose $\psi_j$ so that averages with respect to the $\pi_j$ converge much more quickly than averages with respect to $\pi$ itself.  It is this basic philosophy that we extend in Section \ref{sec:gen} to the computation of  dynamic averages.

\subsection{Averages with Respect to a Given Markov Process}
\label{sec:gen}

The mathematical description of the nonequilibrium umbrella sampling scheme that follows reveals how the stratification strategy developed for the equilibrium case in Section \ref{sec:eus} can be extended to compute nearly arbitrary dynamic statistics. Our interest in this section is computing averages over trajectories of some specified Markov process, $X^{(t)}.$  This process can be time-inhomogenous, i.e., given the value of $X^{(t)},$ the distribution of $X^{(t+1)}$ can depend on the value of $t.$  We compute averages of trajectories evolved up to a first exit time of the process $(t,X^{(t)})$ from a user specified set of times and positions, $D$---i.e.,  trajectories terminate when they first leave the set $D$. We consider averages over trajectories of $X^{(t)}$ run until time
\begin{equation}\label{tau}
\tau= \min\{ t>0:\, (t, X^{(t)}) \notin D\}
\end{equation}
for a set 
$D\in \mathbb{N} \times \mathbb{R}^d$. \jt{In the first numerical example in Section \ref{sec:num}, $D$ is a set of times and positions for which we would like to compute an escape probability. In the second numerical example, $D$ restricts only the times over which we simulate.} The averages are of the form
\begin{equation}\label{f_ave_dynamical}
\mathbf{E}\left[ \sum_{t=0}^{\tau-1} f(t,X^{(t)}) \right].
\end{equation}
We note that the average in \eqref{f_ave_dynamical} is not completely general, in order to streamline the developments below.  Without any modification, we can compute averages similar to \eqref{f_ave_dynamical} but with the argument $(t,X^{(t)})$ in the definitions of $\tau$ and  $f$ replaced by $(t,X^{(t-1)},X^{(t)})$.  On the other hand, expectations with $(t,X^{(t)})$ replaced  by  $(t,X^{(t-m)},\ldots,X^{(t-1)},X^{(t)})$ for $m\geq 2$ cannot be obtained immediately.  These and many more general expectations can, however, be accommodated by applying the algorithm to an enlarged process (e.g., $(t,X^{(t-m)},\ldots,X^{(t-1)},X^{(t)})$) at the cost of storing copies of the enlarged process.  For many expectations, this cost is quite manageable. Finally, we require that $\mathbf{E}\left[ \tau\right] < \infty.$ The limit $\tau\rightarrow \infty$ is considered in Section \ref{sec:erg}.

Below we show that expectations of time-dependent functions can be decomposed as a weighted sum of expectations computed over restricted subsets of the full space and, in turn, how the statistical weights can be computed as expectations over these subsets, mirroring the basic structure of the equilibrium scheme described in Section \ref{sec:eus}. However, as we discuss in Section \ref{sec:fp}, the  algorithm for computing these local expectations departs significantly from the equilibrium case because their form is not known \emph{a priori} in the nonequilibrium setting.

\subsubsection{The Index Process}
\label{sec:index}

\begin{figure*}
\centering
\includegraphics[scale=0.4]{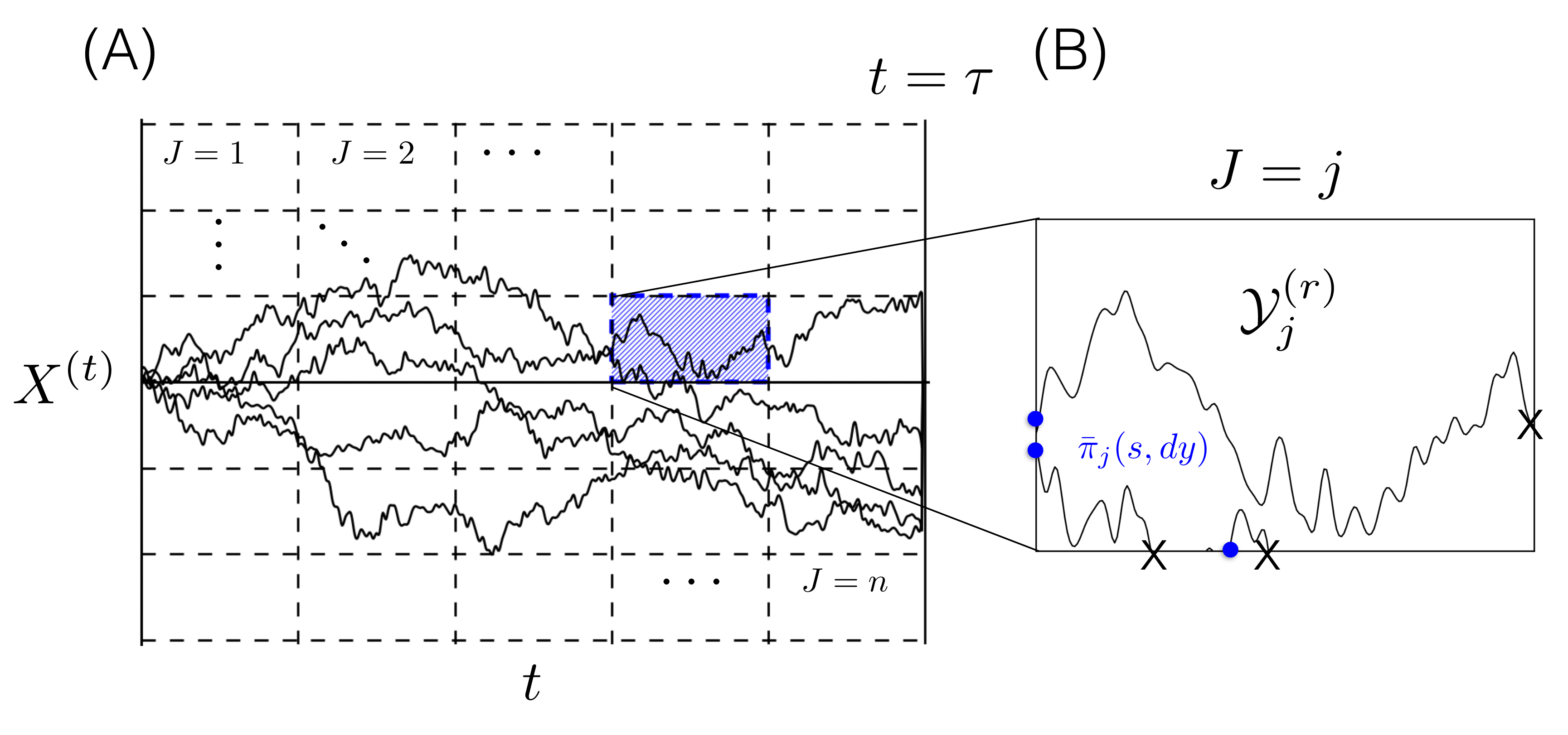}
\caption{Illustration of the stratification of a process $(X^{(t)}, J^{(t)})$ (solid black lines, panel A) via the scheme outlined in Section \ref{sec:gen}. (A) The restricted distributions corresponding to each value of the index process $J^{(t)}$ are outlined as discrete regions of the $(t, X^{(t)})$ space (panel A, black dashed lines). In this depiction, the value of $J^{(t)}$ corresponds to the current cell containing $(t,X^{(t)})$ within a rectangular grid of times and positions. (B) Each of the restricted distributions $\pi_{j}(t,dx)$ are sampled by integrating a locally restricted dynamics $\mathcal{Y}_{j}^{(r)}$ (panel B, black lines). The $\mathcal{Y}_{j}^{(r)}$ process is generated by integrating an excursion of the unbiased process $(X^{(t)}, J^{(t)})$ corresponding to a particular fixed value of $J =j$ (panel A). As each excursion transitions from $J=i$ to $J=j$ with $j \neq i$, the dynamics are stopped and a new excursion is started at a time and point $(s,y)$ (panel B, blue dots) drawn from the flux distribution $\bar \pi_{j}(s,dy)$. 
}\label{neus-illustration}
\end{figure*}

The US scheme in Section \ref{sec:eus} used the basis functions $\psi_j$  to stratify the sampling of the distribution $\pi$ by decomposing averages with respect to $\pi$ into averages with respect to the more easily sampled $\pi_j$.  To arrive at an analogous partitioning of state space for the nonequilibrium case, we introduce an \emph{index process} $J^{(t)}$ that takes values in $\{1,2,\dots,n\}$ and (roughly) labels the point $(t,X^{(t)})$ in time and space, $\mathbb{N}\times \mathbb{R}^d$.  Our objective is to generate fragments of trajectories of $X^{^{(t)}}$ consistent with specific values of $J^{(t)}$ thereby breaking the coupled process $(X^{(t)}, J^{(t)})$ into separate regions corresponding to a given value of $J^{(t)}$ \jt{(see panel A of Figure \ref{neus-illustration})}.

The idea of discretizing a process $X^{(t)}$ according to the value of some user-specified index process is not new in computational statistical mechanics.   For example, in our notation, given a partition of state space $A_1, A_2,\dots, A_n$, the Milestoning procedure \cite{Faradjian2004} and some Markov State Modeling procedures \cite{Schutte2011} correspond to an index process that marks the pairs of sets $(A_i,A_j)$ for $i\neq j$ between which $X^{(t)}$ last transitioned.  In the Milestoning method, the pairs of sets are considered unordered, so that a transition from $A_j$ to $A_i$ immediately following a transition from $A_i$ to $A_j$ does not correspond to a change in $J^{(t)}$, and $J^{(t)}$ can assume \jt{$n= \binom{m}{2}$} distinct values. The original presentation of NEUS on the other hand corresponds to a process $J^{(t)}$ which marks the index of the set $A_j$ containing $X^{(t)}.$   For accurate results, the Milestoning procedure requires that the index process $J^{(t)}$ itself be Markovian.  Even under the best circumstances, that assumption is only expected to hold approximately.  It is not required by the NEUS algorithm.    Our presentation below reveals the full flexibility in the choice of $J^{(t)}$ within NEUS.  That flexibility  is essential in the generalized setting of this article.

 In the developments below we require that $J^{(t)}$ is chosen so that the \emph{joint process $(X^{(t)},J^{(t)})$} is Markovian. This assumption  allows that trajectories can be continued beyond a single transition event (before $\tau$) without additional information about the history of $X^{(t)}$ or $J^{(t)}$.   We do not assume that $J^{(t)}$ alone is Markovian and in general it is not.  Our assumption implies no practical restriction on the underlying Markov process $X^{(t)}$.  When $X^{(t)}$ is non-Markovian, additional variables can often be appended to $X^{(t)}$ to yield a new Markov process to which the developments below can be applied.   A version of this idea is applied in Section \ref{sec:jarzynski}  where we append a variable representing a nonequilibrium work  to an underlying Markov process.
 
\subsubsection{The Eigenproblem}

Given a specific choice of index process $J^{(t)},$ the nonequilibrium umbrella sampling algorithm stratifies  trajectories of $X^{(t)}$ according to their corresponding values of $J^{(t)}$.  That is, for each possible value of the index process, NEUS generates segments of trajectories of $X^{(t)}$ between the times that $J^{(t)}$ transitions to and from $J = j$.  To make this idea more precise, we need to carefully describe the distribution sampled by these trajectory fragments:
\begin{equation}\label{piidyn} \pi_j(t,dx) = \frac{  \mathbf{P}\left[ t < \tau,\, X^{(t)}\in dx,\, J^{(t)} = j\right] }{z_j},
\end{equation}
where 
\begin{equation}\label{zdyn}
z_j = \sum_{t=0}^\infty \mathbf{P}\left[  t < \tau,\, J^{(t)} = j\right].
\end{equation} 
For each $j$, $\pi_j$ is the distribution of time and position pairs $(t,X^{(t)})$ conditioned on $J^{(t)}=j$ and $t < \tau$.  We call the $\pi_{j}$  \emph{restricted distributions}. We have reused the notations $\pi_j$ and $z_j$ from our account of the equilibrium umbrella sampling scheme to emphasize the analogous roles played by those objects in both sections.  Note that here we are treating time as an additional random variable. Also note that in these definitions as well as in the formulas below, $\mathbf{P}$ and $\mathbf{E}$ represent probabilities and expectations with respect to the original, unbiased  $X^{(t)}$ and $J^{(t)}$. We assume that $z_j>0$ for all $j$  since  we can remove the index $j$ from consideration if $z_j=0$.   The $z_j$ are all finite because $\sum_{j=1}^n z_j = \mathbf{E}\left[ \tau\right]$, which we assume is finite.

Observe that
 \begin{align}\label{avefexp2}
 \mathbf{E} \left[ \sum_{t=0}^{\tau-1} f( t,X^{(t)}) \right] &= \sum_{t=0}^\infty \mathbf{E}\left[ f \left(t,X^{(t)} \right),\, t<\tau \right] \notag \\
 &= \sum_{j=1}^n \sum_{t=0}^\infty \int_{x \in \mathbb{R}^d} f(t, x) \notag \\
 & \hspace{0.1cm} \times \mathbf{P}\left[t < \tau,\, X^{(t)}\in dx,\, J^{(t)}=j \right] \notag \\&= \sum_{j=1}^n  z_j\, \langle f\rangle_j, 
\end{align}
where
\begin{equation}\label{fdyn}
\langle f\rangle_j = \sum_{t=0}^\infty \int_{x\in\mathbb{R}^d} f(t,x) \pi_j(t,dx).
\end{equation}
Thus we have a decomposition of \eqref{f_ave_dynamical} analogous to the decomposition of \eqref{fave} in \eqref{avefexp}. Also as in the equilibrium case, the $z_j$ can be computed from averages with respect to the $\pi_j$. To see this, observe that for any $t$  we can  write
\begin{multline}
\sum_{i=1}^n \mathbf{P}\left[t+1< \tau,\, J^{(t+1)}=j,\,J^{(t)}=i\right] \\= \mathbf{P}\left[t+1<\tau,\, J^{(t+1)} = j\right].
\end{multline}
Summing this expression over $t$ we obtain
\begin{multline}\label{summed1}
\sum_{i=1}^n \sum_{t=0}^\infty \mathbf{P}\left[t+1< \tau,\, J^{(t+1)}=j,\,J^{(t)}=i \right] \\= \sum_{t=0}^\infty \mathbf{P}\left[t<\tau,\, J^{(t)} = j\right] - \mathbf{P}\left[ J^{(0)} = j \right].
\end{multline}
These expressions are all bounded by $\mathbf{E}\left[\tau\right]$ and are therefore finite. Expression \eqref{summed1} can be rewritten as an affine eigenequation:
\begin{equation}\label{summed2}
z^\text{\tiny T} G + a^\text{\tiny T} = z^\text{\tiny T},
\end{equation}
where $z$ is defined in \eqref{zdyn},
\begin{equation}\label{Gdyn}
G_{ij} = \frac{\sum_{t=0}^\infty \mathbf{P}\left[t+1< \tau,\, J^{(t+1)}=j,\,J^{(t)}=i \right]} {z_i},
\end{equation}
and
\begin{equation}\label{a}
a_j = \mathbf{P}\left[ J^{(0)} = j \right].
\end{equation}
Equation \eqref{summed2} is the analog of \eqref{eigeq2} in Section \ref{sec:eus}. Here, the matrix element $G_{ij}$ stores the expected number of transitions from  $J=i$ to  $J=j$, normalized by the expected number of time steps with $J=i$. Note that the matrix $G$ is substochastic; that is, it has non-negative entries and rows that sum to a number less than or equal to one. 

To complete the analogy with the umbrella sampling scheme described in Section \ref{sec:eus}, we need to show that the elements of the matrix $G$ are expressible as expectations over the $\pi_j$. Indeed,
\begin{align}
G_{ij} &= \frac{1}{z_i} \int_{x \in \mathbb{R}^d} \sum_{t=0}^\infty\mathbf{P}_{t,x,i}\left[ t+1 < \tau,\, J^{(t+1)}=j \right] \notag \\
& \hspace{0.1cm} \times \mathbf{P}\left[t< \tau,\, X^{(t)}\in dx,\,J^{(t)}=i\right]\notag \\
& = \sum_{t=0}^\infty \int_{x \in \mathbb{R}^d} \mathbf{P}_{t,x,i}\left[t+1 < \tau,\, J^{(t+1)}=j\right]\pi_i(t,dx) \label{Gasave} 
\end{align}
where $\mathbf{P}_{t,x,i}$ is used to denote probabilities with respect to $X$ initialized at time and position $(t,x)$ and conditioned on $J^{(t)}=i$ and $t<\tau.$ Note that in the first line we have appealed to the Markovian assumption on $(X^{(t)},J^{(t)})$.  Had we instead assumed that $J^{(t)}$ alone was Markovian, we could have ignored the $x$ dependence in \eqref{Gasave}. 

Just as for the umbrella sampling algorithm described in Section \ref{sec:eus}, we arrive at a procedure for computing \eqref{f_ave_dynamical} via stratification:
\begin{enumerate}
\item Assemble $G_{ij}$ defined in \eqref{Gdyn} and  $\langle f\rangle_{j} $ defined in \eqref{fdyn} by sampling from the $\pi_j$ defined in \eqref{piidyn}.
\item Solve the affine eigenvector equation \eqref{summed2} for $z$ defined in \eqref{zdyn}.
\item  Compute
\begin{equation}
\mathbf{E}\left[ \sum_{t=0}^{\tau-1}f(t,X^{(t)})\right]  = \sum_{j=1}^n z_j \langle f\rangle_j 
\end{equation}
via \eqref{avefexp2}.
\end{enumerate}

Relative to the scheme in Section \ref{sec:eus}, sampling the restricted distributions $\pi_j$ requires a more complicated procedure.   This is the subject of Section \ref{sec:fp}.   
\jw{In Section \ref{sec:fp}, instead of $G$, we choose to work with the matrix 
\begin{align}\label{barG}
\bar G_{ij}&= 
\frac{\sum_{\ell=0}^\infty \mathbf{P} [S^{(\ell+1)} < \tau, J^{(S^{(\ell+1)})}=j,J^{(S^{(\ell)})}=i ] }{\sum_{\ell=0}^{\infty} \mathbf{P} [J^{(S^{(\ell)})}=j, S^{(\ell)} < \tau]},
\end{align}
where 
\begin{equation}\label{Sl}
S^{(\ell)} =  \min \left \{s>S^{(\ell-1)}:\,J^{(s)}\neq J^{(S^{(\ell-1)})} \right \}
\end{equation}
is the time of the $\ell$th change in the value of $J^{(t)}$ for a given realization of the coupled process $(X^{(t)}, J^{(t)})$.
Likewise, instead of $z$, we choose to work with the weights    
\begin{equation}\label{barz}
\bar z_j = \sum_{\ell=0}^{\infty} \mathbf{P} [J^{(S^{(\ell)})}=j, S^{(\ell)} < \tau].
\end{equation}

We show in Appendix \ref{apx:proofs} that $\bar G$ is related to $G$ by the identity
\begin{equation}\label{eq: G and bar G}
\bar G_{ij} = 
\begin{cases} 
{G_{ij}}/{(1-G_{ii})},& j\neq i\\
0, & j=i,
\end{cases}
\end{equation}
and that $\bar z$ is related to $z$ by 
\begin{equation}\label{eq: z and bar z} 
 \bar z_j =  z_j (1-G_{jj}).
\end{equation}
Therefore, knowledge of $G$ implies knowledge of $\bar G$ and $\bar z$, and the algorithm detailed in the next section could also be expressed in terms of $G$ and $z$ at the cost of additional factors of $1-G_{jj}$ in several formulas.
Moreover, identities \eqref{summed2}, \eqref{eq: G and bar G}, and~\eqref{eq: z and bar z} imply 
 \begin{equation}\label{bareqn}
\bar z^\text{\tiny T} = \bar z^\text{\tiny T} \bar G + a^\text{\tiny T};
\end{equation}
that is, $\bar z$ and $\bar G$ solve the same affine eigenproblem as $z$ and $G$.
We emphasize $\bar G$ and $\bar z$ over $G$ and $z$ only to simplify the presentation and interpretation of the algorithm in Section \ref{sec:fp}.

To give an appealing intuitive interpretation of $\bar G$, we note that for $i \neq j$,  
\begin{multline}\label{netprobflux}
\bar z_i\, \bar G_{ij} = z_i\, G_{ij} \\= {\sum_{\ell=0}^\infty \mathbf{P}\left[S^{(\ell+1)}<\tau,\, J^{(S^{(\ell+1)})}=j,\,J^{(S^{(\ell)})}=i \right]}.
\end{multline}
We  refer to this quantity as the \emph{net probability flux} from $J=i$ to $J=j$; it 
is the expected number of transitions of the process $J^{(t)}$ from $J=i$ to $J=j$ before time $\tau$.  
The matrix $\bar G$ stores the relative probabilities of transitions to different values of $J$ before time $\tau$ and $\bar z_j$ is the expected number of transitions into $J=j$ before time $\tau$.
}

Finally, we remark that rapid convergence of the scheme in practice rests upon the choice of $J^{(t)}$. Roughly, one should choose the index process  so that the variations in estimates of  the required averages with respect to the $\pi_j$ (e.g., estimates of the $G_{ij}$) are small.  In practice, this requires that transitions between values of $J^{(t)}$ are frequent, which is the analog of selecting the biases in equilibrium US to limit the range of the free energy over each subset of state space (see \cite{Thiede2016, Dinner2016}). In Section \ref{sec:num} we describe this and other important implementation details in the context of particular applications.

\section{A General NEUS Fixed-Point Iteration}
\label{sec:fp}

In this section we present a detailed algorithm for computing \eqref{f_ave_dynamical} by the stratification approach outlined in Section \ref{sec:gen}. To accomplish this one must be able to generate samples from the restricted distributions $\pi_{j}(t, dx)$. 
In NEUS, the restricted distributions are sampled by introducing a set of Markov processes 
\begin{equation}
 \mathcal{Y}_j^{(r)} = \left ( T_j^{(r)}, Y_j^{(r)}, I_j^{(r)} \right )
\end{equation}
called \emph{excursions} whose values are triples of a time $T_j^{(r)}$, a position $Y_j^{(r)}$, and a value of the index process $I_j^{(r)}$.
To avoid confusion, we consistently use the variable $r$ for the time associated with an excursion $\mathcal{Y}_j^{(r)}$ and the variable $t$ for the time associated with the process $(t, X^{(t)}, J^{(t)})$.

Roughly speaking, each excursion is a finite segment of a trajectory of the process $(t,X^{(t)}, J^{(t)})$ with $J=j$.
These segments are stopped either on reaching time $\tau$ or at the first time when $J \neq j$.
To be precise, excursions are generated as follows:
\begin{enumerate}
 \item Draw an initial time and position pair $(T_j^{(0)}, Y_j^{(0)})$ from the distribution $\bar \pi_j(s,dy)$ specified below or from an estimate of that distribution. Set $\mathcal{Y}_j^{(0)} = (T_j^{(0)}, Y_j^{(0)}, j)$.
 \item Set $T^{(r+1)}_j = T^{(r)}_j+1$, and generate $(Y_j^{(r+1)}, I_j^{(r+1)})$ from the distribution of $(X^{(T_j^{(r+1)})},J^{(T_j^{(r+1)})})$ conditioned on $X^{(T_j^{(r)})}=Y^{(r)}_j$ and $J^{(T_j^{(r)})} = j$. 
 \item Stop on reaching time $\tau$ or when $J \neq j$. That is, stop when $r$ reaches
 \begin{equation}\label{eq:length of excursion}
  \rho_j = \min \left \{r\geq 0 : I_j^{(r)} \neq j\text{ or }(T_j^{(r)},Y_j^{(r)})\notin D\right \}.
 \end{equation}
 
\end{enumerate}
The excursions $\mathcal{Y}_{j}^{(r)}$ are illustrated in Figure \ref{neus-illustration} for a particular choice of index process. 

For the excursions $\mathcal{Y}_j^{(r)}$ to sample the restricted distribution $\pi_j(t,dx)$,  we must take the initial distribution $\bar \pi_j(s,dy)$ to be the distribution of times $s$  and positions $y$ at which the process   $(t, X^{(t)}, J^{(t)})$ transitions from a state $J^{(s-1)}=i$ with $i\neq j$ to state $J^{(s)}=j$ (see Section \ref{sec:flux_distributions} and Appendix \ref{apx:excursions sample restricted dists}). We call these distributions the \emph{flux distributions}. 

In general, the flux distributions $\bar \pi_j(s,dy)$ are not known \emph{a priori} and must be computed approximately.
In the NEUS algorithm, we begin with estimates of the flux distributions and the matrix $\bar G$. 
We then compute excursions initialized from these estimates of the flux distributions. 
From the excursions and the current estimate of $\bar G$, we compute statistics which are used to improve the estimates of both the flux distributions and $\bar G$.
Thus, NEUS is an iteration designed to produce successively better estimates of the flux distributions and $\bar G$ simultaneously. 

In Section \ref{sec:fixed_point}, we derive a fixed-point equation solved by $\bar G$ and the flux distributions, and we motivate NEUS as a self-consistent iteration for solving this equation.
In Section \ref{sec:stoc}, we describe the complete NEUS algorithm in detail and interpret it as a stochastic approximation algorithm \cite{Kushner2003} for solving the fixed-point equation derived in Section \ref{sec:fixed_point}. \jt{In the Supplementary Material, we analyze a simple four-site Markov model to clearly illustrate the structure of this self-consistent iteration and the terminology of the framework.}

\subsection{The Flux Distributions}
\label{sec:flux_distributions}

Before deriving the fixed-point problem and the corresponding stochastic approximation algorithm, we define the flux distributions $\bar \pi_{j}(s,dy)$ precisely. 
We let
\begin{multline}\label{barpi}
\bar \pi_j(s,dy) \\ \hspace{0.2cm} = \frac {\sum_{\ell=0}^\infty {\mathbf{P}\left[S^{(\ell)}=s,\,s < \tau,\, X^{(s)}\in dy,\, J^{(s)}=j\right]}}{\bar z_j}
\end{multline}
be the distribution of time and position pairs $(S^{(\ell)}, X^{(S^{(\ell)})})$ conditioned on $J^{(S^{(\ell)})} = j$.
With this definition of $\bar \pi_j(s,dy)$, an excursion $\mathcal{Y}^{(r)}_j$ samples the  restricted distribution $\pi_j(t,dx)$ in the sense that 
\begin{align}\label{pi_from_pi_bar}
  &\pi_{j}(t,dx) \notag \\
  &=  
  \frac{\bar z_j}{z_j} 
  {\bf P} 
  \left [ t <\rho_j+T_j^{(0)}, Y_j^{(t-T_j^{(0)})} \in dx\right ]
  \nonumber \\
  &= \frac{\bar z_j}{z_j} \displaystyle\sum_{s=0}^{t} \int_{y\in \mathbb{R}^d}	{\bf P}_{s,y,j} \left[ t < \sigma(s)\wedge \tau, X^{(t)} \in dx \right] \bar{\pi}_{j}(s, dy),
\end{align}
where 
\begin{equation}
 \sigma(s) = \min\{r>s:\, J^{(r)}\neq J^{(s)}\}
\end{equation}
and $\rho_j$ is defined in~\eqref{eq:length of excursion}. 
We prove~\eqref{pi_from_pi_bar} in Appendix \ref{apx:excursions sample restricted dists}.

Given~\eqref{pi_from_pi_bar}, we may express any average over $\pi_j$  as an average over $\bar \pi_j$.  
For example,
\begin{equation}
\bar G_{ij}  =  \sum_{s=0}^\infty \int_{y \in \mathbb{R}^d} \mathbf{P}_{s,y,i}\left[ J^{(\sigma(s))}=j,\, \sigma(s)<\tau \right]\bar\pi_i(s,dy). \label{Hasbarave} 
\end{equation}
Moreover, from \eqref{avefexp2}, we can express general averages as
\begin{equation}
 \mathbf{E} \left[ \sum_{t=0}^{\tau-1} f( t,X^{(t)}) \right] = \sum_{j=1}^n  \bar z_j\, \bar {\langle f\rangle}_j, 
\end{equation}
where
\begin{equation}\label{fdynbar}
\begin{split}
\bar{ \langle f\rangle}_j &=\displaystyle \sum_{s=0}^{\infty} \int_{y\in \mathbb{R}^d} \sum_{t=s}^\infty \int_{x\in\mathbb{R}^d} f(t,x)\, \\& \hspace{0.1cm} \times	{\bf P}_{s,y,j} \left[ t < \sigma(s)\wedge \tau, X^{(t)} \in dx \right] \bar{\pi}_{j}(s, dy).
\end{split}
\end{equation}
We use these facts in our interpretation of the NEUS algorithm in Section \ref{sec:fixed_point}.

Instead of working directly with the flux distributions, we find it convenient to express both the fixed-point problem and the algorithm in terms of the probability distribution of time and position pairs $(t,X^{(t)})$ conditioned on observing a transition from $J=i$ to $J=j$ at time $t$, i.e., in terms of 
\begin{align}\label{gamma1}
&\gamma_{ij}(s, dy) \nonumber \\&= \frac{1}{\bar z_i \bar G_{ij}} \nonumber\\& \times \sum_{\ell=0}^\infty 
{\bf P} \left [ s = S^{(l+1)} < \tau, J^{(S^{(l)})}=i, J^{(s)}=j, X^{(s)} \in dy \right ]
\nonumber \\
&= \frac{1}{\bar G_{ij}} \sum_{r=0}^{\infty} \int_{w\in \mathbb{R}^d}{\bf P}_{r,w,i} [s=\sigma(r),\,s<\tau,\,   X^{(s)} \in dy,\, J^{(s)} = j]\,\nonumber\\& \hspace{0.1cm} \times \bar\pi_i ( r, dw)
\end{align}
which is defined only for $s>0.$  To simplify notation, we let $\gamma$ denote the set of all conditional distributions $\gamma_{ij}$. 
\jt{Recall from \eqref{netprobflux} that $\bar{z}_i \bar{G}_{ij}$ is the net probability flux from $J = i$ to $J= j$.} The following simple but key identity relates $\gamma$ to the flux distributions $\bar \pi_j$:
\begin{equation}\label{gamma2barpi}
\bar \pi_j (s, dy) = \frac{1}{\bar z_j}\begin{cases} \sum_{i\neq j} \bar z_i\,\bar G_{ij}\, \gamma_{ij}(s,dy), &\text{if } s>0\\
 a_j \,{\bf P}[X^{(0)} \in dy \mid J^{(0)} = j] & \text{if } s=0.
 \end{cases}
\end{equation}
The $s > 0$ term is the contribution from transitions into state $J =j$ from the neighboring state $J=i$, and the $s=0$ term accounts for the initial $t=0$ contribution of the underlying process when $J=j$. We emphasize that both the fixed-point problem and the iteration that we define below could be expressed in terms of the flux distributions $\bar \pi_j$ instead of $\gamma$. We choose to express them in terms of $\gamma$ because the resulting formalism more naturally captures the implementation of the method used to generate our numerical results in Section \ref{sec:num}.

\subsection{The Fixed-Point Problem}
\label{sec:fixed_point}

We now derive the fixed-point problem. Our goal is to find an expression of the form
\begin{equation}\label{fpeq}
\left( \mathcal{G}( \bar G,  \gamma), \Gamma( \bar G,  \gamma )\right)= \left(\bar G,\gamma\right)
\end{equation}
that characterizes the desired matrix ${\bar G}$ and collection of probability measures $\gamma$ as the fixed-point of a pair of maps $\mathcal{G}(\tilde {G}, \tilde \gamma)$ and $\Gamma(\tilde G, \tilde \gamma)$ that take as arguments  approximations $\tilde G$ of $\bar G$ and $\tilde \gamma$ of $\gamma$ and return, respectively, a new substochastic matrix and a new collection of probability measures.

To this end, we define a function mapping $\tilde G$ and $\tilde \gamma$ to an approximation of the flux distribution $\bar\pi_j.$   We denote this function by the corresponding capital letter $\bar {{\Pi}}_j.$   
Based on  \eqref{bareqn} and \eqref{gamma2barpi}, we define
\begin{align}\label{barPi}
	\bar {{\Pi}}_j  &(s, dy; \tilde G, \tilde \gamma) \notag \\& =\frac{1}{\tilde z_j}\begin{cases}
	{\sum_{i\neq j} \tilde z_i\,\tilde G_{ij}\, \tilde \gamma_{ij}(s,dy)} &\text{if } {s>0}, \\
	 a_j \,{\bf P}[X^{(0)} \in dy \mid J^{(0)} = j] &\text{if }  {s=0},
	 \end{cases}
	 \end{align}
where $\tilde z$ solves the equation $\tilde z^\text{\tiny T} = \tilde z^\text{\tiny T} \tilde G + a^\text{\tiny T}$. \jt{The matrices $\tilde G$ that we consider are strictly substochastic. We assume that $\tilde G$ is also irreducible, in which case the solution $\tilde z$ exists and is unique.} To motivate the definition above, we observe that for the exact values $\bar G$ and $\gamma$, $\bar \pi_j(s,dy) = \bar{\Pi}_j(s,dy; \bar G, \gamma)$ by~\eqref{gamma2barpi}.
Moreover, given $\tilde G$ and samples from $\tilde \gamma$, one can generate  samples from $\bar {{\Pi}}_j (s, dy; \tilde G, \tilde \gamma)$; see Section \ref{sec:stoc}.
This is crucial in developing a practical algorithm to solve the fixed-point problem.

At this point we are ready to  define the functions $\mathcal{G}$ and $\Gamma$ appearing in \eqref{fpeq} above. For a substochastic matrix $\tilde G$ and a collection of probability distributions $\tilde \gamma = \{\tilde \gamma_{ij}\}$, define the substochastic matrix
\begin{multline}\label{Hfunc}
	\mathcal{G}_{ij}(\tilde G, \tilde \gamma)\\=  \sum_{s=0}^\infty \int_{y \in \mathbb{R}^d} \mathbf{P}_{s,y,i}\left[ J^{(\sigma(s))}=j,\, \sigma(s)<\tau \right]\,\bar \Pi_i(s, dy; \tilde G, \tilde \gamma)
\end{multline}
and the collection of probability distributions
\begin{widetext}
\begin{equation}\label{gamma2}
	 \Gamma_{ij} (s, dy; \tilde G, \tilde \gamma) \propto  \sum_{r=0}^{\infty} \int_{w\in \mathbb{R}^d} {\bf P}_{r,w,i} [s=\sigma(r),\,s<\tau,\,   X^{(s)} \in dy,\, J^{(s)} = j]\, \bar \Pi_i ( r, dw; \tilde G, \tilde \gamma).
\end{equation}
\end{widetext}
Because $\bar \Pi_j(\bar G,\gamma) = \bar \pi_j,$ expressions \eqref{Hasbarave} and \eqref{gamma1} imply 
that $\mathcal{G}(\bar G, \gamma) = \bar G$ and $\Gamma_{ij}( \bar G, \gamma) = \gamma_{ij},$ establishing 
our fixed-point relation \eqref{fpeq}.

Having fully specified the fixed-point problem, we can now consider iterative methods for its solution. One approach would be to fix some $\varepsilon \in (0,1]$ and compute the  deterministic fixed-point iteration
\begin{equation}
\begin{split}
  \tilde G(m+1) &=  \tilde G(m) + \varepsilon \left ( \mathcal{G}(\tilde G(m),\tilde \gamma(m)) - \tilde G(m) \right ),
  \text{ and } \\
 \tilde \gamma(m+1) &= \tilde \gamma(m) + \varepsilon \left ( \Gamma(\tilde G(m), \tilde \gamma(m)) - \tilde \gamma(m) \right ) ,
 \end{split}
 \label{Hdfpit}
\end{equation}
given initial guesses $\tilde G(0)$ and $\tilde \gamma(0)$ for $G$ and $\gamma$, respectively. One would typically choose $\varepsilon =1$ in this deterministic iteration; we consider arbitrary $\varepsilon \in (0,1]$ to motivate the stochastic approximation algorithm developed in Section \ref{sec:stoc}.

\jt{In practice, computing $\mathcal{G}$ and $\Gamma$ in the right hand side of \eqref{Hdfpit} requires computing averages over trajectories of $(X^{(t)},J^{(t)})$ initiated from $\bar \Pi_j(\tilde G(m), \tilde \gamma(m))$.  While we cannot hope to compute these integrals exactly, we can construct a stochastic algorithm  approximating  the iteration in \eqref{Hdfpit} using a finite number of sampled trajectories. The resulting scheme, which we detail in  Section \ref{sec:stoc}, fits within the basic stochastic approximation framework.   }

\subsection{A Stochastic Approximation}
\label{sec:stoc}

In this section, we present the full NEUS algorithm and we interpret it as a stochastic approximation algorithm analogous to the deterministic fixed-point iteration~\eqref{Hdfpit}. 
In NEUS, as in the fixed-point iteration, we generate a sequence of approximations $\tilde G(m)$ and $\tilde \gamma(m)$, converging  to $\bar G$ and $\gamma$, respectively.   
During the $m$th \jt{iteration} of the NEUS algorithm, 
we update the current approximations $\tilde G(m)$ and $\tilde \gamma(m)$ based on statistics gathered from $K$ independent  excursions $\mathcal{Y}_j^{(r)}(m)=(T^{(r)}_j, Y^{(r)}_j, I_j^{(r)})$ defined according to the rules governing $\mathcal{Y}_j^{(r)}$ enumerated above with $(T^{(0)}_j,Y^{(0)}_j)$ drawn from $\bar \Pi_j(\tilde G(m), \tilde \gamma(m))$, the current (at the $m$th iteration of the scheme) estimate of the  flux distribution $\bar \pi_j$.

We now state the NEUS algorithm. To simplify the expressions below, we sometimes omit the iteration number $m$. The algorithm proceeds as follows:\\
\begin{widetext}
\nopagebreak
\begin{enumerate}
\item Choose initial approximations $\tilde G(0)$ and $\tilde \gamma(0)$ of $\bar G$ and $\gamma$, respectively. Fix the number $K$ of independent excursions $\mathcal{Y}_j^{(r)}(m)$ to compute for each restricted distribution $\pi_j(t, dx)$. Choose the maximum number of new points $L$ included in the update to the empirical approximations of the distributions $\tilde \gamma_{ij}(m)$.\\

\item For each $j=1,2,\dots,n$ generate $K$ independent excursions  
\begin{equation}
\mathcal{Y}_{ik}^{(r)}=(T^{(r)}_{ik}, Y^{(r)}_{ik}, I_{ik}^{(r)}) \text{ for } k=1,2,\dots, K.
\end{equation}
Let 
  \begin{equation}
  \rho_{ik} = \min \left \{r\geq 0 : I_{ik}^{(r)} \neq j\text{ or }(T_{ik}^{(r)},Y_{ik}^{(r)})\notin D\right \}
 \end{equation} 
 be the length of the excursion $\mathcal{Y}_{ik}^{(r)}$ as in~\eqref{eq:length of excursion}.\\
\item Let 
\begin{equation}
M_{ij}(m) =  \sum_{k=1}^{K} \mathbf{1}_{\{j\}}( I_{ik}^{(\rho_{ik})})\,
\mathbf{1}_{D}( T^{(\rho_{ik})}_{ik},Y^{(\rho_{ik})}_{ik})
\end{equation}
be the number of $i$ to $j$ transitions of the index process observed while generating  the excursions $\mathcal{Y}_{ik}^{(r)}(m)$.
 Let $\left \{T_{ij}^{(\ell)} \right \}_{\ell=1}^{M_{ij}(m)}$
  and  $\left \{Y_{ij}^{(\ell)} \right \}_{\ell=1}^{M_{ij}(m)}$
 be the times $T^{(\rho_{ik})}_{ik}$ and positions $Y^{(\rho_{ik})}_{ik}$ for which $I_{ik}^{(\rho_{ik})}=j$ and $Y_{ik}^{(\rho_{ik})} \in D$.  \\

\item  Compute
\begin{equation}
\hat G_{ij}(m) = \frac{M_{ij}(m)}{K},
\end{equation}
\begin{equation} \label{gamma_update}
	\hat \gamma_{ij}(s, dy; m) =  	\begin{cases}
	\frac{1}{L \wedge M_{ij}(m)}
	\sum_{\ell=1}^{L \wedge M_{ij}(m)} \mathbf{1}_{T_{ij}^{(\ell)}}(s)\,
	\delta_{Y_{ij}^{(\ell)}}(dy)
	&\text{ if }
	M_{ij}(m)>0, \\
	0 &\text{ if } M_{ij}(m)=0,
	\end{cases}
\end{equation}
and
\begin{equation}
\hat{\langle f\rangle }_i(m) = \frac{1}{K}\sum_{k=1}^{K} \sum_{r=0}^{\rho_{jk}-1} f \left (T_{jk}^{(r)}(m),Y_{jk}^{(r)}(m) \right ), 
\end{equation}
where $L \wedge M_{ij}(m) = \min\{ L, M_{ij} \jt{(m)}\}$. In Equation \eqref{gamma_update}, $\delta_x$ represents the Dirac delta function centered at position $x.$\\
 
\item Replace the deterministic iteration \eqref{Hdfpit} by the approximation
\begin{equation}\label{Gup}
\tilde G_{ij}(m+1) = \tilde G_{ij}(m)  + \varepsilon_m \left(\hat G_{ij}(m)-\tilde G_{ij}(m)\right)
\end{equation}
and 
\begin{equation}
\tilde \gamma_{ij}(m+1) \\= \tilde \gamma_{ij}(m) + {\varepsilon_m }\left(\hat{\gamma}_{ij}(m) - \tilde \gamma_{ij}(m)\right)\left( \frac{\mathbf{1}_{\{M_{ij}(m)>0\}}}{ I_{ij}(m)}\right)
\end{equation}
where 
\begin{equation}
I_{ij}(m) = \frac{1}{m+1}\sum_{\ell=0}^{m}\mathbf{1}_{\{M_{ij}(\ell)>0\}}
\end{equation}
and $\varepsilon_m>0$ satisfies
\begin{equation}\label{eq:conditions on epsilon_m}
\sum_{m=1}^\infty \varepsilon_m = \infty \quad\text{and}\quad \sum_{m=1}^\infty \varepsilon_m^2 < \infty.
\end{equation}\\

\item Update the expectations
 \begin{equation}\label{fup}
 \tilde {\langle f \rangle}_i (m+1) = \tilde {\langle f \rangle}_i (m) + \varepsilon_m\left(  \hat{\langle f\rangle }_i(m) -  \tilde {\langle f \rangle}_i (m)\right).
 \end{equation}\\ 
 
 \item Once the desired level of convergence has been reached, compute
 \begin{equation}
 \mathbf{E}\left[ \sum_{t=0}^{\tau-1} f(t, X^{(t)})\right] \approx \sum_{j=1}^n \tilde z_j(m)\, \tilde {\langle f \rangle}_i (m),
 \end{equation}
where the vector 
$\tilde z(m)$ solves $\tilde z^{\text{\tiny T}}(m) = \tilde z^{\text{\tiny T}}(m)\, \tilde G(m) + a^\text{\tiny T}.$\\
\end{enumerate}
\end{widetext}

We now interpret NEUS as a stochastic approximation algorithm analogous to the deterministic fixed-point iteration~\eqref{Hdfpit}.
First, we observe that $\hat G(m)$ approximates $\mathcal{G}(\tilde G(m), \tilde \gamma(m))$ in the following sense.
Suppose we were to compute a sequence 
$ \hat G(n), \hat G(n+1), \dots,  \jt{\hat G(n+k-1)}$
as in NEUS, except holding the values of $\tilde G(n)$ and $\tilde \gamma(n)$ fixed.  We would then have  that $\mathbf{E}\left[ \hat G(n+i)\right] = \mathcal{G}(\tilde G(n), \tilde \gamma(n)),$ and that each of the  $\hat G(n+i)$ were independent (conditionally on $\tilde G(n)$ and $\tilde \gamma(n)$).  A Law of Large Numbers would therefore apply and we could conclude that   
\begin{equation}\label{eq:gdot approx property}
 \lim_{k \rightarrow \infty} \frac{1}{k} \sum_{i=0}^{k-1}  \jt{ \hat G(n+i)} = \mathcal{G}(\tilde G(n), \tilde \gamma(n)) .
\end{equation}
 The distribution $ \gamma_{ij}(m)$ approximates $\Gamma_{ij}(\tilde G(m), \tilde \gamma(m))$ in a similar sense.
Therefore, the NEUS iteration~\eqref{Gup} is a version of the deterministic fixed-point iteration~\eqref{Hdfpit} but with a shrinking sequence $\varepsilon_m$ instead of a fixed $\varepsilon$ and with random approximations instead of the exact values of $\mathcal{G}$ and $\Gamma$.
The conditions~\eqref{eq:conditions on epsilon_m} on the sequence $\varepsilon_m$ are common to most stochastic approximation algorithms~\cite{Kushner2003}; they ensure convergence of the iteration when $\mathcal{G}$ and $\Gamma$ can only be approximated up to random errors. 

We remark that in practice the empirical measures $\tilde \gamma(m)$ are stored as lists of time and position pairs. The update in \eqref{Gup} allows the number of pairs stored in these lists to grow with each iteration.  This can lead to impractical memory requirements for the method. We therefore limit the size of each list $\tilde \gamma_{ij}(m)$ to a fixed maximum value by implementing a selection step in which the points that have been stored for the most iterations are removed to make room for the points in the updates of $\tilde \gamma_{ij}(m)$ when this maximum is exceeded. Also, in our numerical experiments in Section \ref{sec:num}, we use  $\varepsilon_m = 1/(m+1)$ in which case, 
 \begin{equation}
 \tilde G_{ij}(m) = \frac{1}{m+1} \sum_{\ell=0}^{m} \hat{G}_{ij}(\ell)\quad\
\end{equation}
 \text{and}
 \begin{equation}
 \quad
 \tilde \gamma_{ij}(m) = \frac{1}{\sum_{\ell=0}^{m}\mathbf{1}_{\{M_{ij}(\ell)>0\}}} \sum_{\ell=0}^{m}  \hat{\gamma}_{ij}(\ell).
 \end{equation}
This and other details of our implementation are explained in Section \ref{sec:num}. 

\jw{The implementation detailed above borrows ideas from several earlier modifications of the basic NEUS algorithm.  The use of a linear system solve for the weights $z$ was introduced in  \cite{Vanden-Eijnden2009}.  In the scheme presented above, the number of samples, $K$, of the process $Y_{j}^{(r)}$ is fixed at the beginning of each iteration of the scheme. In this aspect, the implementation above is similar to the Exact Milestoning approach presented in \cite{Bello-Rivas2015}.   With the number of samples of $Y_j^{(r)}$ fixed, the total amount of computational effort, as measured in number of time steps of the process $X^{(t)}$, becomes a random variable (with expectation $K {\bf E} [\sigma(S^{(\ell)})]$).  In practical applications, it may be advantageous to fix the total computational effort expended per iteration in each $J=j$. An alternative version of the NEUS scheme is therefore to fix the total computational effort expended (or similarly the number of numerical integration steps) and allow the number of samples, $K$, to be a random number.  In our tests (not shown here), neither implementation showed a clear advantage provided that a sufficient number of samples, $K$, was generated to compute the necessary transition statistics.  

It is also important to note that if the number of points used in the representation of $\tilde \gamma$ is restricted (as it typically has to be in practice), any of the implementations of NEUS that we have described has a systematic error that decreases as the number of points increases \emph{or} as the work per iteration increases. Earlier implementations of NEUS \cite{Warmflash2007, Dickson2009-1, Dickson2009-2, Vanden-Eijnden2009, Dickson2011} computed transition statistics that were normalized with respect to the simulation time spent associated with each $J=j$ rather than the number of samples of $Y_j^{(r)}$ generated.  This implementation choice leads to a scheme with a systematic error that vanishes only as the number of points allowed in the representation of $\tilde \gamma$ grows, regardless of the work performed per iteration.}

\section{Ergodic Averages}
\label{sec:erg}

In this section we consider the calculation of ergodic averages with respect to a general (not necessarily time-homogenous) Markov process. We also describe the simplifications that occur when the target Markov process is time-homogenous as in the original NEUS algorithm. 

\jt{ In order to ensure that the definitions in this section are sensible,} we require that
\begin{equation}
\lim_{\tau \rightarrow \infty} \frac{1}{\tau}\sum_{t=0}^{\tau-1} \mathbf{P}\left[X^{(t)}\in dx,\, J^{(t)}=i \right]
\end{equation}
exists as a probability distribution on $\mathbb{R}^d \times \{1,2,\dots,n\}$
and let
\begin{equation}
\pi(dx) = \lim_{\tau \rightarrow \infty} \frac{1}{\tau}\sum_{t=0}^{\tau-1} \mathbf{P}\left[ X^{(t)}\in dx\right].
\end{equation}
This general ergodicity requirement allows processes $X^{(t)}$ with periodicities or time dependent forcing.

Our goal is to compute ergodic averages of the form
\begin{equation}
\lim_{\tau\rightarrow \infty} \frac{1}{\tau}\sum_{t=0}^{\tau-1} \mathbf{E}\left[ f(X^{(t)})\right] = \int_{x\in \mathbb{R}^d} f(x)\pi(dx).
\end{equation}
\jt{ To that end, we fix a deterministic time horizon $\tau>0$ in \eqref{zdyn} and \eqref{Gdyn}; the condition $t<\tau$ can thus be written as an upper bound of $\tau-1$ on the summation index.} If we divide both sides of \eqref{summed2} by $\tau$ and take the limit $\tau \rightarrow \infty,$ we obtain the equation 
\begin{equation}\label{stateig}
  z^\text{\tiny T}   G =  z^\text{\tiny T}
  \end{equation}
   where now
\begin{equation}\label{zerg}
 z_j = \lim_{\tau\rightarrow \infty} \frac{1}{\tau}\sum_{t=0}^{\tau-1} \mathbf{P}\left[J^{(t)}= j\right].
\end{equation}
and
\begin{equation}\label{Ferg}
G_{ij} = \lim_{\tau\rightarrow \infty}\frac{  \sum_{t=0}^{\tau-2} \mathbf{P}\left[ J^{(t+1)}=j,\,J^{(t)}=i\right]}
{ \sum_{t=0}^{\tau-1} \mathbf{P}\left[ J^{(t)}=i\right]}.
\end{equation}
Note that the matrix $ G$ is now stochastic and that
$
\sum_{j=1}^n  z_j = 1.
$
We can   rewrite the ergodic average of $f$ as
\begin{equation}
\int_{x\in \mathbb{R}^d} f(x) \pi(dx) = \sum_{j=1}^n z_j\, \langle f\rangle_j,
\end{equation}
where
\begin{equation}\label{ferg}
\langle f\rangle_j =   \int_{x\in\mathbb{R}^d} f(x) \pi_j(dx)
\end{equation}
and we  represent the large $\tau$ limit of the position marginal distribution of $\pi_j$ defined in \eqref{piidyn} as
\begin{equation}
\pi_j(dx) = \lim_{\tau\rightarrow \infty }
\sum_{t=0}^{\tau-1}
  \pi_j(t,dx).
   \end{equation}
These formulas indicate that the only modification of the algorithm in Section \ref{sec:fp} that is required to compute a long-time average is to set $\tau = \infty$ in the definition of the processes $Y_i(\tilde G, \tilde \gamma)$, to set $a=0$  in \eqref{barPi}, and let $\tilde z$ solve $ \tilde z^\text{\tiny T} =  \tilde z^\text{\tiny T} \tilde G$ with $\sum_{j=1}^n \tilde z_j = 1.$ In other words, the algorithm seamlessly transitions from solving the initial value problem to solving the infinite time problem as $\tau$ becomes large.

When the joint process $(X^{(t)},J^{(t)})$ is time-homogenous and stationary and our goal is to compute the average of a position dependent observable $f(x)$  with respect to the stationary distribution $\pi$ of $X^{(t)},$ the above  relations can be further simplified.  In this case, 
\begin{equation}
\pi_j(dx)   = \frac{1}{z_j}\lim_{t\rightarrow \infty} \mathbf{P}\left[X^{(t)}\in dx,\,J^{(t)}=j\right],
\end{equation}
where $z_j$ defined in \eqref{zerg} becomes
\begin{equation}
z_j = \lim_{t\rightarrow \infty } \mathbf{P}\left[ J^{(t)} = j\right].
\end{equation}
The matrix $G$ in  \eqref{Ferg} can now be written
\begin{equation}
G_{ij} = \lim_{t\rightarrow\infty } \mathbf{P}\left[ J^{(t+1)}=j\,|\, J^{(t)}=i\right]
\end{equation}
and the vector $\langle f\rangle_j$ defined in \eqref{ferg} becomes
\begin{equation}
\langle f\rangle_j = \int_{x\in \mathbb{R}^d} f(x) \pi_j(dx).
\end{equation}
These simplifications lead to a version of the original NEUS method \cite{Warmflash2007} that employs a direct method for solving for the weights similar to the scheme in \cite{Vanden-Eijnden2009}.

In \cite{Vanden-Eijnden2009} and \cite{Dickson2009-2}   the basic NEUS approach was extended to the estimation of transition rates between sets for a stationary Markov process.  Implicit in this extension was the observation  that any  algorithm that can efficiently compute averages with respect to the stationary distribution of a time-homogenous Markov process can be applied to computing dynamic averages more generally by an enlargement of the state space, i.e., by applying the scheme to computing stationary averages for a higher dimensional time-homogenous Markov process.   
This idea is also central to  Exact Milestoning  \cite{Bello-Rivas2015}, which extends the original Milestoning procedure \cite{Faradjian2004} to compute steady-state averages with respect to a time-homogenous Markov process and is very similar in structure to steady-state versions of NEUS.

\section{Numerical Examples}
\label{sec:num}

Here we illustrate the flexibility of the generalized algorithm with respect to both the means of restricting the trajectories (the choice of the $J^{(t)}$ process) and the averages that can be calculated.  Specifically, in Section \ref{sec:j_prcs} we discuss our choice of the $J^{(t)}$ process.  In Section \ref{sec:dipeptide} we show how finite-time hitting probabilities can be calculated by discretizing the state space according to both time and space. In Section \ref{sec:jarzynski} we show how free energies can be obtained by discretizing the state space according to time and the irreversible work.

\subsection{One Choice of the $J^{(t)}$ Process}
\label{sec:j_prcs}

Rapid convergence of the scheme outlined in Section \ref{sec:fp} rests on the choice of $J^{(t)}.$  Perhaps the most intuitive choice is
\begin{equation}\label{Jpart}
J^{(t)} = \sum_{j=1}^n j \mathbf{1}_{A_j}(t,X^{(t)})	
\end{equation}
where the subsets $A_1,A_2,\dots,A_n$ partition $\mathbb{N}\times \mathbb{R}^d$.  Indeed, earlier steady-state NEUS implementations \cite{Warmflash2007, Dickson2009-1, Dickson2009-2, Vanden-Eijnden2009, Dickson2011} employed an analogous rule using a partition of the space variable (the time variable was not stored or partitioned). However, even with an optimal choice of the subsets $A_1,A_2,\dots,A_n$, \eqref{Jpart} has an important disadvantage:  in many situations, $X^{(t)}$ frequently recrosses the boundary between neighboring subsets $A_i$ and $A_j$, which slows convergence.  Fortunately, there are many alternative choices of $J^{(t)}$ that approximate the choice in \eqref{Jpart} while mitigating this  issue.  We give one simple and intuitive alternative which we use in the numerical examples that follow. 

Let $\psi_j$ be a set of non-negative functions on $\mathbb{N}\times \mathbb{R}^d$ for which $\sum_{j=1}^n\psi_j = 1$.  The $\psi_j$  are generalizations of the functions $\mathbf{1}_{A_j}$ in that they serve to restrict trajectories to regions of state space. In practice, given a partition of space   $A_1,A_2,\dots,A_n,$ the $\psi_j$ can be chosen to be smoothed approximations of the functions $\mathbf{1}_{A_j}$. Given a trajectory of $X^{(t)}$, the rule defining $J^{(t)}$ is as follows. Initially, choose $J^{(0)}\in \{1,2,\dots,n\}$ with probabilities proportional to $\{\psi_1(0,X^{(0)}),\psi_2(0,X^{(0)}),\dots,\psi_n(0,X^{(0)})\}.$  At later times $J^{(t)}$ evolves according to the rule\\

\begin{enumerate}\label{Jrule}
\item If $\psi_{J^{(t-1)}}(t,X^{(t)})> 0$ then $J^{(t)} = J^{(t-1)}.$\\
\item  Otherwise sample $J^{(t)}$ independently from $\{1,2,\dots,n\}$ according to  probabilities  $\{\psi_1(t,X^{(t)}),\psi_2(t,X^{(t)}),\dots,\psi_n(t,X^{(t)})\}.$\\
\end{enumerate}
\noindent 
While transitions out of $J^{(t)}=i$ occur when $X^{(t)}$ leaves the support of $\psi_i,$ transitions back into $J^{(t)}=i$ can only occur outside of the support of $\psi_j.$  Thus, this transition rule allows one to separate in space the values of $X^{(t)}$ at which $J^{(t)}$ transitions away from $i$ from those where $J^{(t)}$ transitions into $i,$ mitigating the recrossing issues mentioned above.  
  
 In our examples, we discretize time and only one additional ``collective variable'' (a dihedral angle in Section \ref{sec:dipeptide} and the nonequilibrium work in Section \ref{sec:jarzynski}). Here we denote the collective variable by $\phi$, and we discretize it within some interval of values $[a,b]$ (though it may take values outside this interval).   In both examples $[a,b]$ is evenly discretized into a set of points $\{a+k(b-a)/m_\phi\}_{k=0}^{m_\phi}$ for  some integer $m_\phi$.  Letting $\phi_j$ be any of the points in that discretization, we set 
\begin{widetext}
\begin{align}\label{psi_def}
\psi_{j}(t,x) \propto \begin{cases} \left[1 - \frac{1}{\Delta_\phi} |\phi(x)  \phi_j|\right]\mathbf{1}_{[a,b]} & \text{if }  |\phi(x) - \phi_j | \leq \Delta_\phi \text{ and } t \in [t_{start}^{j}, t_{end}^{j})\\
0& \text{otherwise }
\end{cases}
\end{align}
\end{widetext}
where $\Delta_\phi$ is some fixed value controlling the width of the support of $\psi_j$, and the indicator $\mathbf{1}_{[a,b]}$ restricts the terminal functions. Recall that the $\psi_j$ are required to sum to 1.  We choose $t_{start}^j$ and $t_{end}^j$ to equally divide the interval $[0,\tau)$, where, in our examples,  $\tau$   is a fixed time horizon. The function $\psi_j$ is largest when $t\in [t_{start}^{j}, t_{end}^{j})$ and $\phi(x) = \phi_j$.  The supports of the various $\psi_j$ correspond to products of overlapping intervals in the $\phi$ variable, but non-overlapping intervals in time.  The fact that $\psi_j$ depends on time is essential in our examples. 

\subsection{Finite-Time Hitting Probability}
\label{sec:dipeptide}

In this section we compute the probability, $P_{BA}(\tau_{\text{max}})$, of hitting a set $B$ before a separate set $A$ and before a fixed time $\tau_{\text{max}} > 0$ given that the system is at a point $X^{(0)} \notin A \cup B$ at time $t=0$. In the case where $X^{(0)}$ and $B$ are separated by a large free energy barrier while $X^{(0)}$ and $A$ are not, computing $P_{BA}(\tau_{\text{max}})$ can be challenging since trajectories that contribute to $P_{BA}(\tau_{\text{max}})$ are rare in direct simulations. To compute $P_{BA}(\tau_{\text{max}})$ via the scheme in Section \ref{sec:stoc}, we let the stopping time $\tau$ be the minimum of $\tau_{\text{max}}$ and the first time, $t,$ at which $X^{(t-1)}$ is in either $A$ or $B,$ i.e.,  $\tau-1 = \min\{\tau_{A}, \tau_{B} , \tau_{\text{max}}-1\}$ where $\tau_{A}$ and $\tau_{B}$ are the first times that $X^{(t)}$ enters  the sets $A$ and $B$ respectively. Strictly speaking, to write $\tau$ in the form in \eqref{tau}, we need to replace $(t,X^{(t)})$ in that equation by $(t,X^{(t-1)},X^{(t)}).$  The set $D$ corresponding to our choice of $\tau$ is then $D = \{(t,x,y):\, t< \tau_\text{max},\, x\notin (A\cup B)\}$.  As we have already mentioned, this can be done without further modification of the scheme. Then  $f(t,X^{(t)})$ in \eqref{f_ave_dynamical} is 
\begin{equation}
	f(t, X^{(t)}) = \mathbf{1}_{B}(X^{(t)}) .
\end{equation} 

\begin{figure}
\centering
\includegraphics[scale=0.6]{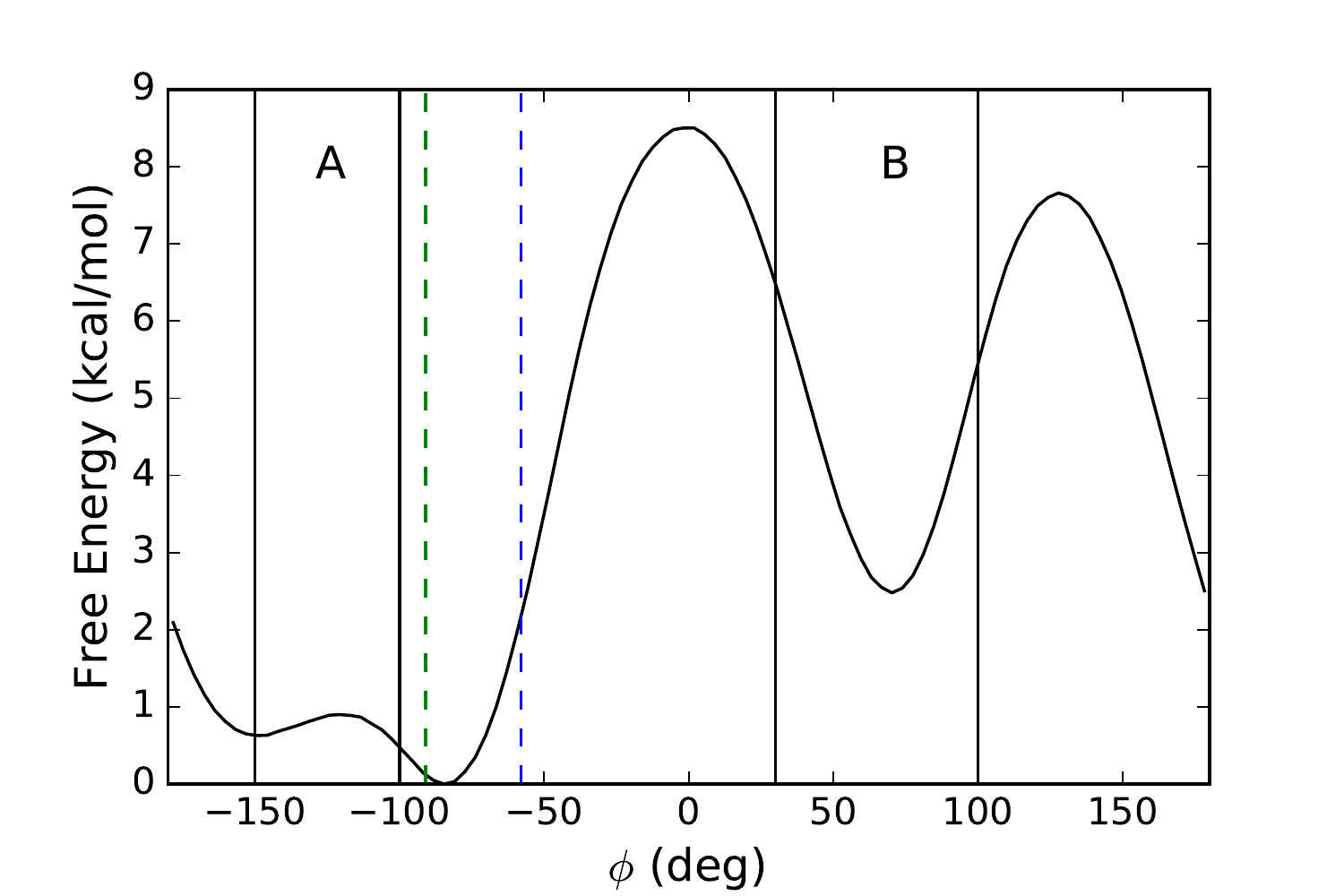}
\caption{Free energy (black curve) of the alanine dipeptide projected onto the $\phi$ dihedral angle, with sets $A$ and $B$ indicated. The initial positions of $X^{(0)}$ at $\phi = -58.0\degree$ (blue) and $\phi = -91.0\degree$ (green) are shown as vertical dashed lines. The free energy is computed from the method presented in Section \ref{sec:eus} as implemented in \cite{Thiede2016}.}\label{sets_on_phi}
\end{figure}

The system that we simulate is the alanine dipeptide (CH$_3$-CONH-C$^\alpha$H(C$^\beta$H$_3$)-CONH-CH$_3$) in vacuum modeled by the CHARMM 22 force field \cite{MacKerell1998}. We use the default Langevin integrator \cite{Schneider1978} implemented in LAMMPS  \cite{Plimpton1995}, with a temperature of 310 K, a timestep of 1 fs and a damping coefficient of $30\,\text{ps}^{-1}$. The SHAKE algorithm is used to constrain all bonds to hydrogens \cite{Ryckaert1977}. We consider the system to be in set $A$ if $-150\degree< \phi < -100\degree$  and in set $B$ if $30\degree < \phi < 100\degree$ (Figure \ref{sets_on_phi}). We  discretize  time into intervals of $t_{end} - t_{start} = 10^3$  time steps with a terminal time of $\tau_{\text{max}} =  10^4$ time steps.  We use the rule outlined in Section \ref{sec:j_prcs} for the evolution of $J^{(t)}$ with the $\psi_j$ of the form in \eqref{psi_def}.  The $\phi_j$ in \eqref{psi_def} are chosen from the set $\{-100\degree, -74\degree, -48\degree, -22\degree, 4\degree, 30\degree\}$ with $[a,b] = [-100\degree, 30\degree]$ and $\Delta_\phi=20\degree$.  

We generate the initial point $X^{(0)}$ by running an unbiased simulation at 310 K and choosing a single point $X^{(0)}$ between the sets $A$ and $B$. The vector $a$ defined in \eqref{a} is 
\begin{equation} \label{a_def}
a_{j} = \frac{\psi_{j} (0,X^{(0)}) }{\sum_{i=1}^{n} \psi_{i} (0,X^{(0)})}.
\end{equation}
Note that the initial condition at $J^{(0)}$ can be drawn from an ensemble of configurations with minimal changes to the algorithm, but we restrict our attention to the initial condition consisting of a single point. To evaluate the performance of the algorithm in Section \ref{sec:stoc}, we choose two points from our direct simulation, one at $\phi = -58.0\degree$ and one at $\phi = -91.0\degree$. The former is chosen to allow the NEUS results to be compared with results from unbiased direct simulations, while the latter provides a more challenging test because $P_{BA}$ becomes small when $X^{(0)}$ is close to $A$. 

We set $K = 100$ and $L = 1$ and perform a total of $10^4$ iterations (about 7.2 $\mu$s of dynamics) of the scheme in Section \ref{sec:stoc} for each starting point. Each step of the process $\mathcal{Y}_{j}^{(r)}(\tilde{G}(m), \tilde{\gamma}(m))$ corresponds to $10$ time steps  of the physical model. The $\tilde \gamma_{ij}$ are represented as lists of time and position pairs with associated weights. We cap the maximum size of those lists at 25 entries. If $\tilde \gamma_{ij}$ reaches this maximum size, each new entry overwrites the oldest previous entry.  Because the lists are empty at the start of the calculation, we restrict sampling in Step 2 of the algorithm in Section \ref{sec:stoc} to regions with at least one stored entry point (``progressive initialization'' in \cite{Dickson2011}).  When required, a sample $(S,Y)$ is drawn from $\bar{\Pi}_{j}(s,dy; \tilde{G}(m), \tilde{\gamma}(m))$ by the following. With probability $a_{j} / z_{j}$, set $S=0$ and select $Y$ from $\mathbf{P} [X^{(0)} \in dy | J^{(0)} = j]$, or with the remaining probability select an index $I$ proportional to the flux $\tilde{z}_{i}\,\tilde{G}_{ij}$ and then select $(S,Y)$  from the list of weighted samples comprising $\tilde{\gamma}_{Ij} (m).$  For each $j$ we compute  $\langle f\rangle_j = P^{j}_{BA} = M_{jB} /(mK)$ where $M_{jB}$ is the total number of transition events of $X^{(r)}_j$ into $B$ observed after $m$ iterations ($mK$ is the total number of excursions in state $j$ after $m$ iterations). The estimate of $P_{BA}(\tau_{\text{max}})$ after $m$ iterations is then computed as $P_{BA}(\tau_{\text{max}}) = \sum_{j=1}^{n} P^{j}_{BA}\, \tilde{z}_{j}(m)$.  
 
\begin{figure}
\centering
\includegraphics[scale=0.6]{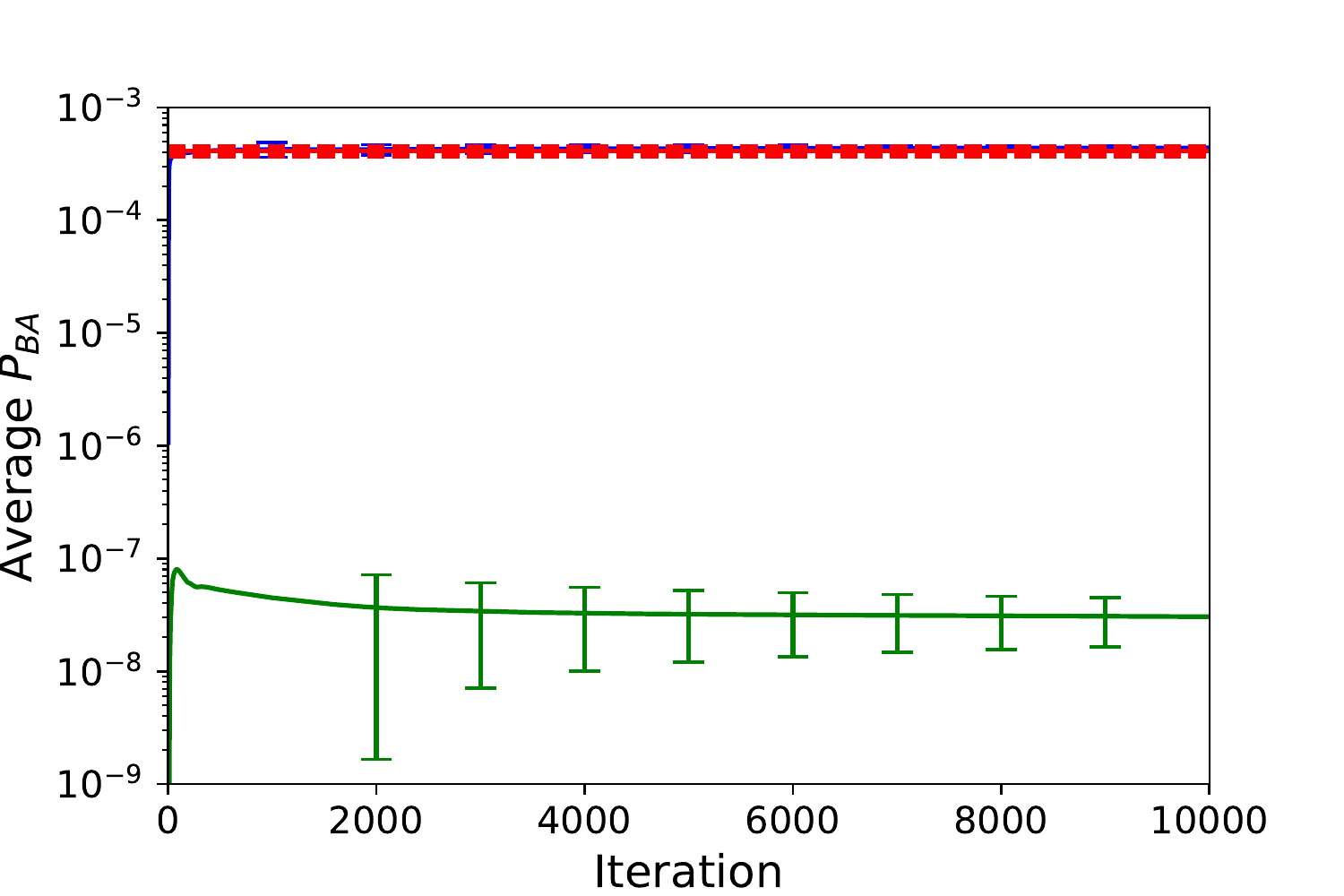}
\includegraphics[scale=0.6]{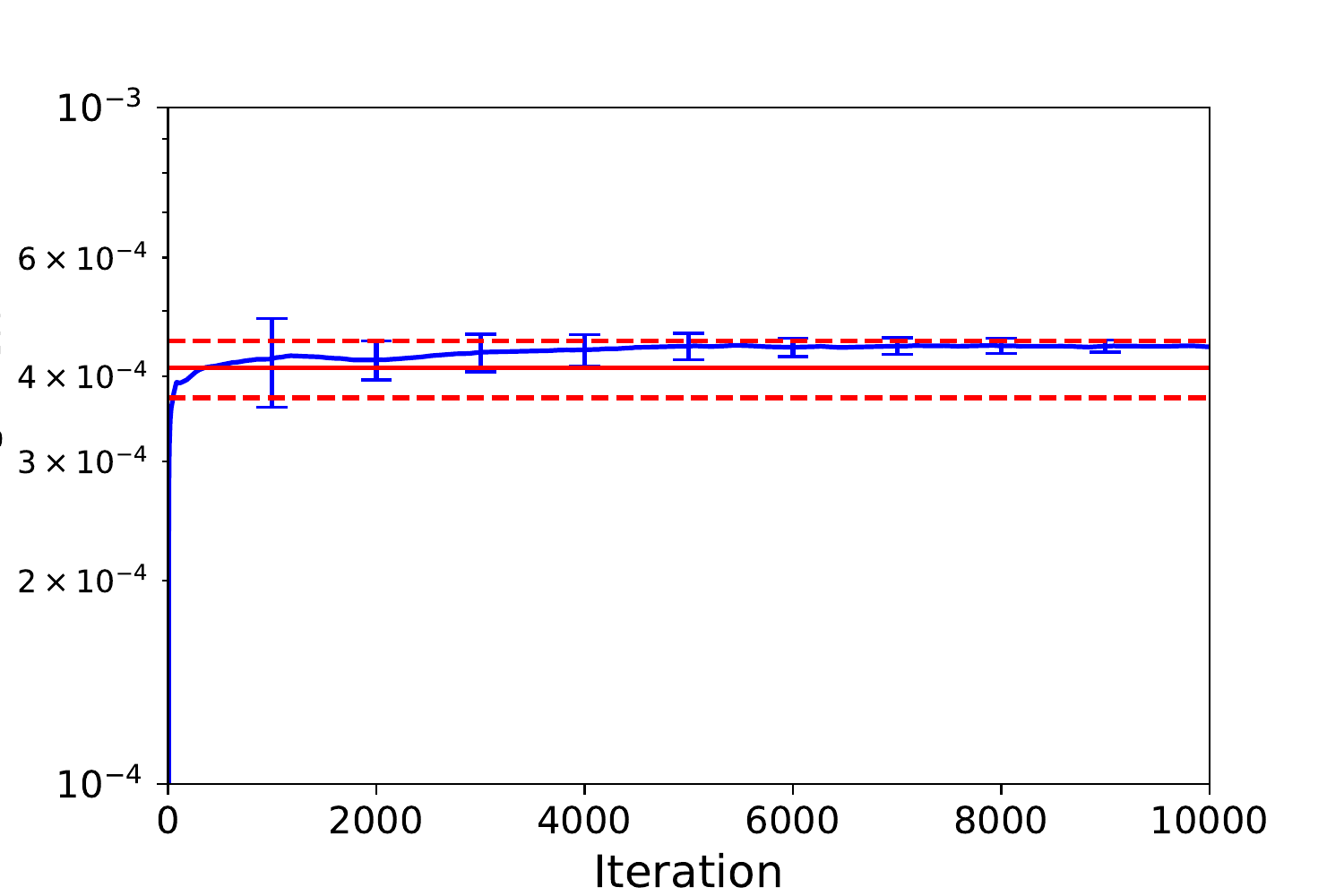}
\caption{Running estimate of $P_{BA}$ from NEUS for dynamics starting at $\phi=-58.0\degree$ (blue, upper curve; error bars are computed every 1000 iterations and indicate $\pm 2.262 s / \sqrt{n}$ where $s$ is the standard error estimated from $n=10$ independent NEUS simulations) compared to the final result from direct simulation (red solid line; dashed lines indicate $\pm1.96\sqrt{p(1-p)/n}$, where $n=10^6$ is the number of physically weighted trajectories generated and $p$ is the estimate of $P_{BA}$ from the direct simulation). Also shown is the estimate from NEUS for dynamics starting from $\phi=-91.0\degree$ (green, lower curve; error bars computed similarly as the blue curve). The estimate at each iteration is computed as the average of the previous 1000 iterations. Lower panel is a magnification of the upper panel.}\label{pba}
\end{figure}

\begin{figure}
\centering
\includegraphics[scale=0.6]{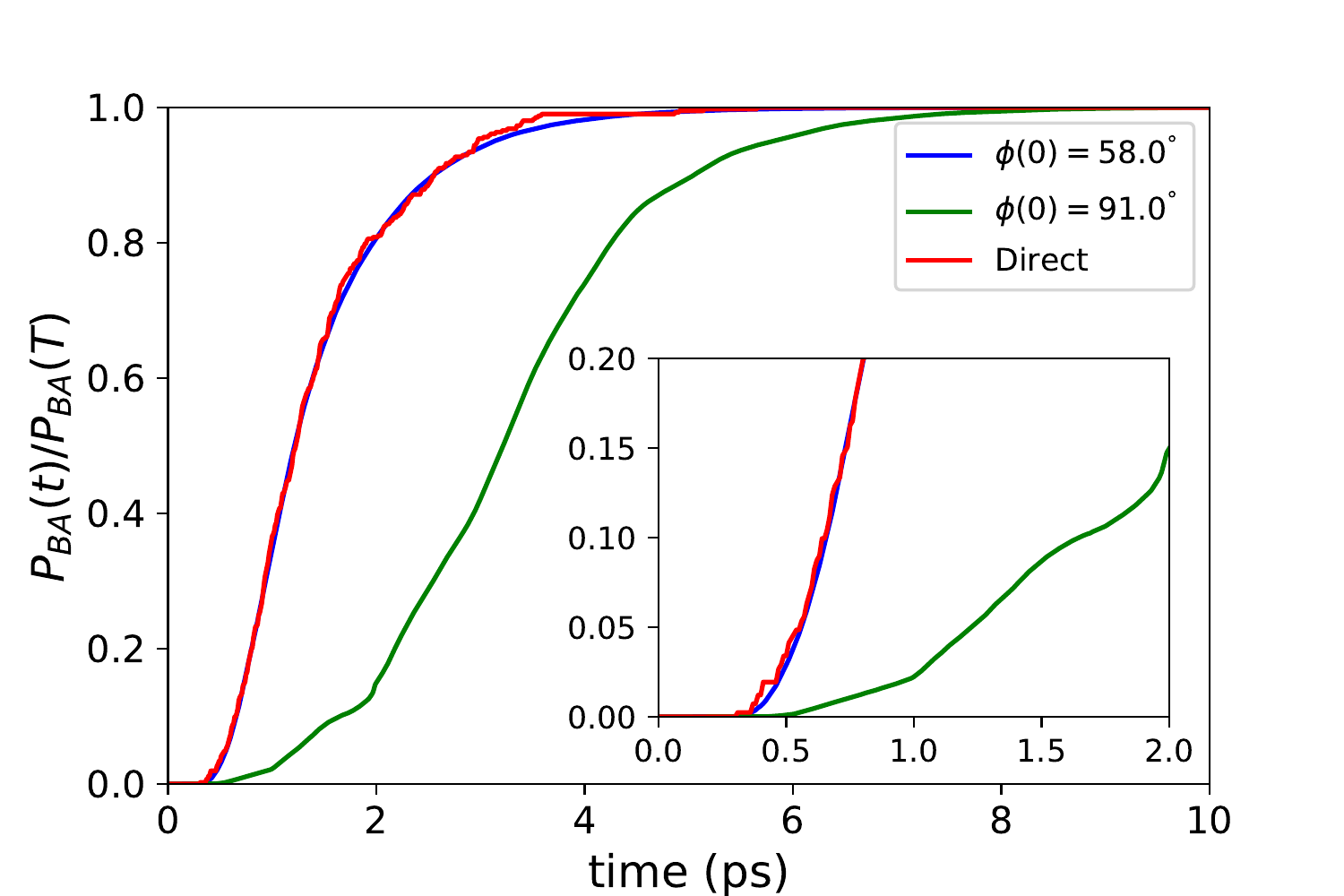}
\caption{Estimate of the cumulative distribution function of the time to enter set $B$ conditioned on not entering $A$ from NEUS for the dynamics starting at $\phi = -58.0\degree$ (blue) and $\phi = -91.0\degree$ (green) compared to the result from the direct simulation (red). (Inset) The early time portion is shown. The estimate from each NEUS simulation at each time is computed as an average over the last 1000 iterations of the calculation and then averaged over 10 independent NEUS simulations.}
\label{p_BA_of_t}
\end{figure}

To assess  the efficiency of the trajectory stratification, we also estimate $P_{BA}(\tau_{\text{max}})$ by  integrating an ensemble of $n=10^6$ unbiased dynamics trajectories for $\tau_{\text{max}}$ time steps from the initial point $X^{(0)}$. In this case,  $P_{BA}(\tau_{\text{max}}) \approx N_B / N $, where $N_{B}$ is the number of trajectories that hit set $B$ before set $A.$ To assess the accuracy of the NEUS result, we perform 10 independent NEUS calculations. In each NEUS simulation, we estimate the value of $P_{BA}$ as the average over the final 1000 iterations of each simulation and compute the mean of this estimate over 10 independent NEUS simulations. We obtain $P_{BA}(\tau_{\text{max}}) \approx 4.43 \times 10^{-4}$ from NEUS and $P_{BA}(\tau_{\text{max}}) \approx 4.12 \times 10^{-4}$ from direct simulation for the starting point at $\phi = -58.0\degree$ (Figure \ref{pba}). In this case, the NEUS result is within the 95\% confidence interval [$3.72 \times 10^{-4}$, $4.52 \times 10^{-4}$] (estimated as $\pm1.96\sqrt{p(1-p)/n}$, where $p$ is the estimate of $P_{BA}$ from the direct simulation) for the direct simulation estimate given the number of samples. We obtain  $P_{BA}(\tau_{\text{max}}) \approx 2.78 \times 10^{-8}$ from NEUS for the starting point at $\phi = -91.0\degree$, consistent with the fact that none of the unbiased trajectories reached $B$ before $A$ in this case. From the same data (for either NEUS or direct simulation), one can easily assemble estimates of $P_{BA}(t)$ for any $t\leq \tau_{\text{max}}$ by counting only those transitions into $B$ that occur before $t$ time steps.    Up to a normalization, $P_{BA}(t)$ is the cumulative distribution function for the time that it takes $X^{(t)}$ to enter $B$ conditioned on not entering $A$.  Estimates of this cumulative distribution function compiled from the NEUS and  direct simulation data are plotted in  Figure \ref{p_BA_of_t}. The NEUS results show excellent agreement with the results from the direct simulation.

\begin{figure*}
\centering
\includegraphics[scale=0.57]{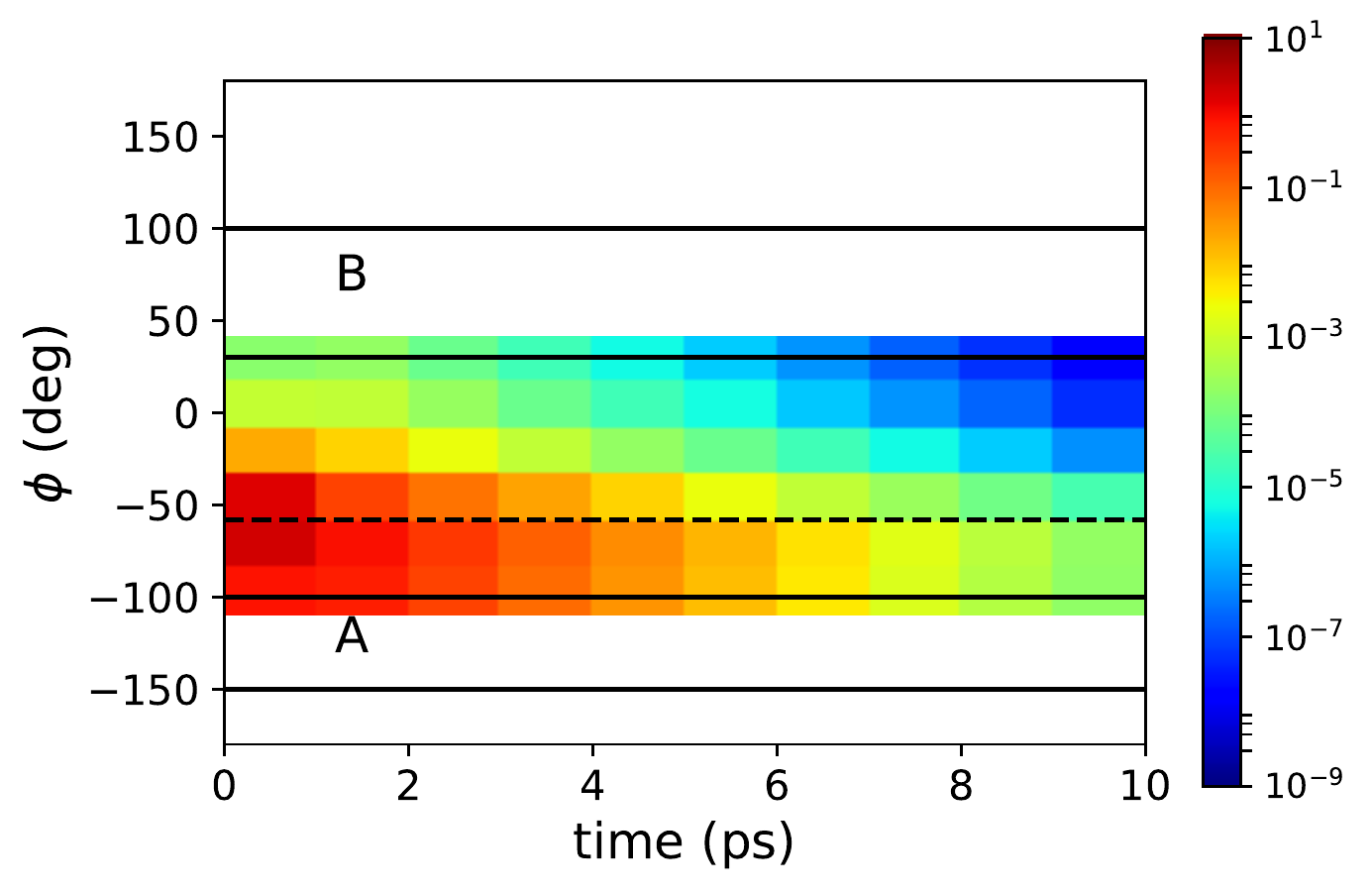} 
\includegraphics[scale=0.57]{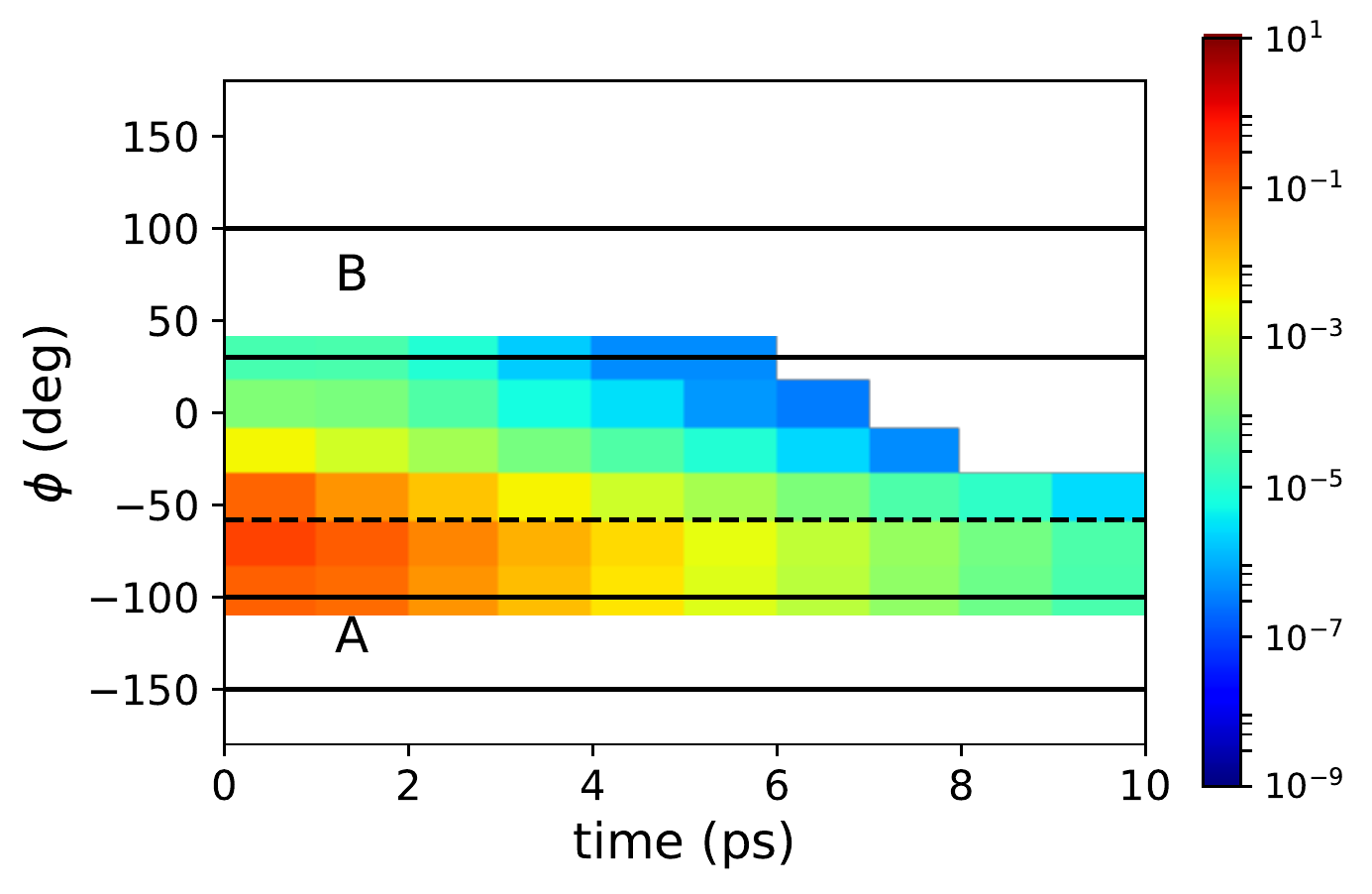}
\includegraphics[scale=0.57]{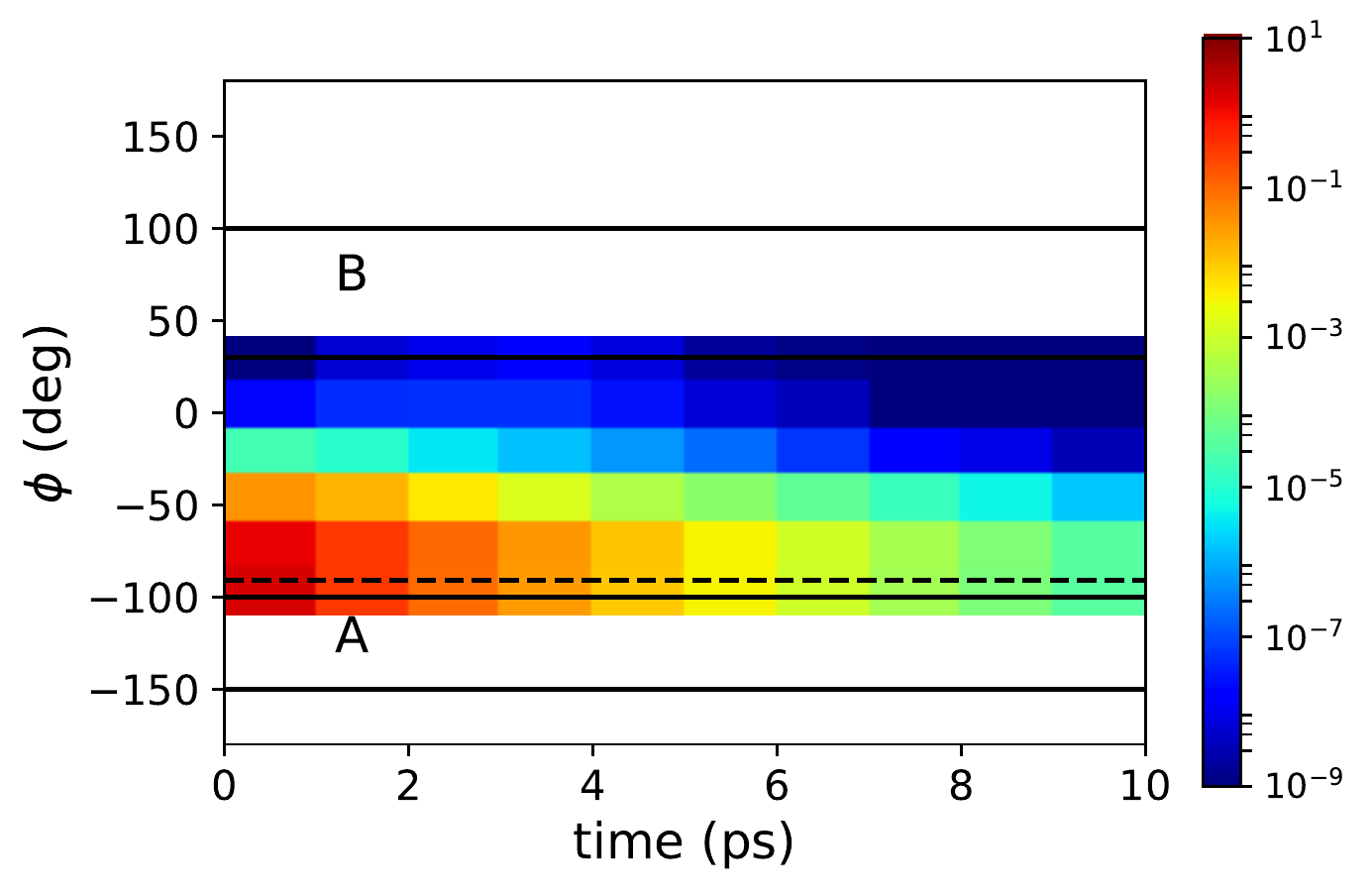}
\includegraphics[scale=0.57]{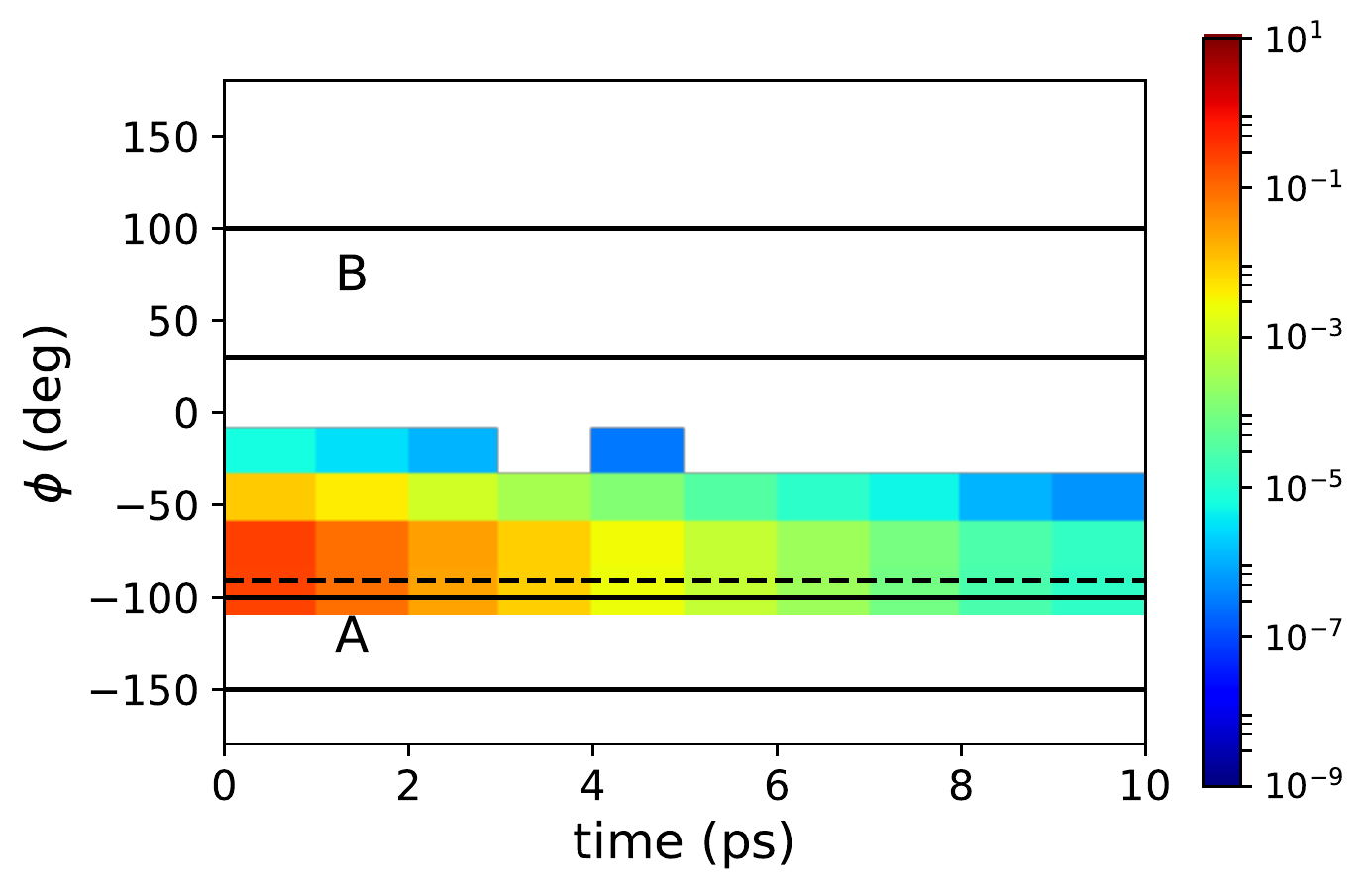}
\caption{Estimates of the subset weights from NEUS (left) and direct simulations (right).  Upper panels show the dynamics starting from $\phi = -58.0\degree$ (dashed line) and lower panels show the dynamics starting from $\phi = -91.0 \degree$ (dashed line). White space represents subsets which were not sampled.}
\label{weights} 
\end{figure*}

Spatiotemporal plots of the weights computed from the converged NEUS calculations and the direct simulations are shown in Figure \ref{weights}. For both starting points, the stratification scheme is able to efficiently sample events with weights spanning 12 orders of magnitude. When $X^{(0)}$ is close to the boundary of set $A$, accurate estimation  of the very small probability $P_{BA}( \tau_{\text{max}})$  depends sensitively on the ability to realize a set of very rare trajectories, ruling out the use of direct simulation.
	
\subsection{Free Energy Differences via the Jarzynski Equation}
\label{sec:jarzynski}

In this section, we show how a specific choice of the $J^{(t)}$ process enables us to stratify a path-dependent variable, specifically, the accumulated  work appearing in the Jarzynski equation \cite{Jarzynski1997, Lelievre2010}. For a statistical model defined by a density proportional to $\exp[-V(x)]$ (e.g., $V(x)$ is a potential function or a log-likelihood), the normalization constant is $Q = \int e^{-V(x)}dx$. In fields ranging from statistics to chemistry, a ratio of normalization constants is often used to compare models \cite{Chipot2007, KassRaftery:1995:bayesfactors}. Subject to certain conditions \cite{Jarzynski1997, Neal2001}, the Jarzynski equation relates the ratio of normalization constants to an average over paths of a time-dependent process, $X^{(t)}$:
\begin{equation}\label{jarzynski_rel}
 \frac{Q_t}{Q_0} =  \mathbf{E} \left[ \exp(- W^{(t)}) \right]
\end{equation}
where 
\begin{equation}
W^{(t)} = \sum_{\ell=0}^{t-1} V(\ell+1,X^{(\ell)}) - V(\ell,X^{(\ell)}),\qquad W^{(0)}=0	
\end{equation}
and we refer to $\Delta F = -\log(Q_{t} / Q_{0})$ as the free energy difference. For example, for a small time discretization parameter, $dt,$ a suitable choice of dynamics is
\begin{equation}\label{XforJar}
	X^{(t+1)} = X^{(t)} -  \frac{\partial V(t+1, X^{(t)})}{\partial x}\,dt + \sqrt{2\,dt}\, \xi_{t}
\end{equation}
where  $\xi_{t}$ is a standard Gaussian random variable and $X^{(0)}$ is drawn from $p_0 \propto \exp[-V(0,x)]$. 

Formula \eqref{jarzynski_rel} suggests a numerical procedure for estimating free energy differences in which one simulates many  trajectories of $X^{(t)}$, evaluates the work $W^{(t)}$ for each, and then uses this sample to compute the expectation on the right hand side of \eqref{jarzynski_rel} approximately.  This approach has been particularly useful in the context of single-molecule laboratory experiments \cite{Hummer2001, Hummer2003}. A well-known weakness of this strategy in the fast-switching (small $t$) regime is large statistical errors result from the fact that low-work trajectories  contribute significantly to the expectation but are infrequently sampled \cite{Hummer2001, Ytreberg2004, Oberhofer2005, Jarzynski2006, Vaikuntanathan2011}.

\begin{figure}
\centering
\includegraphics[scale=0.6]{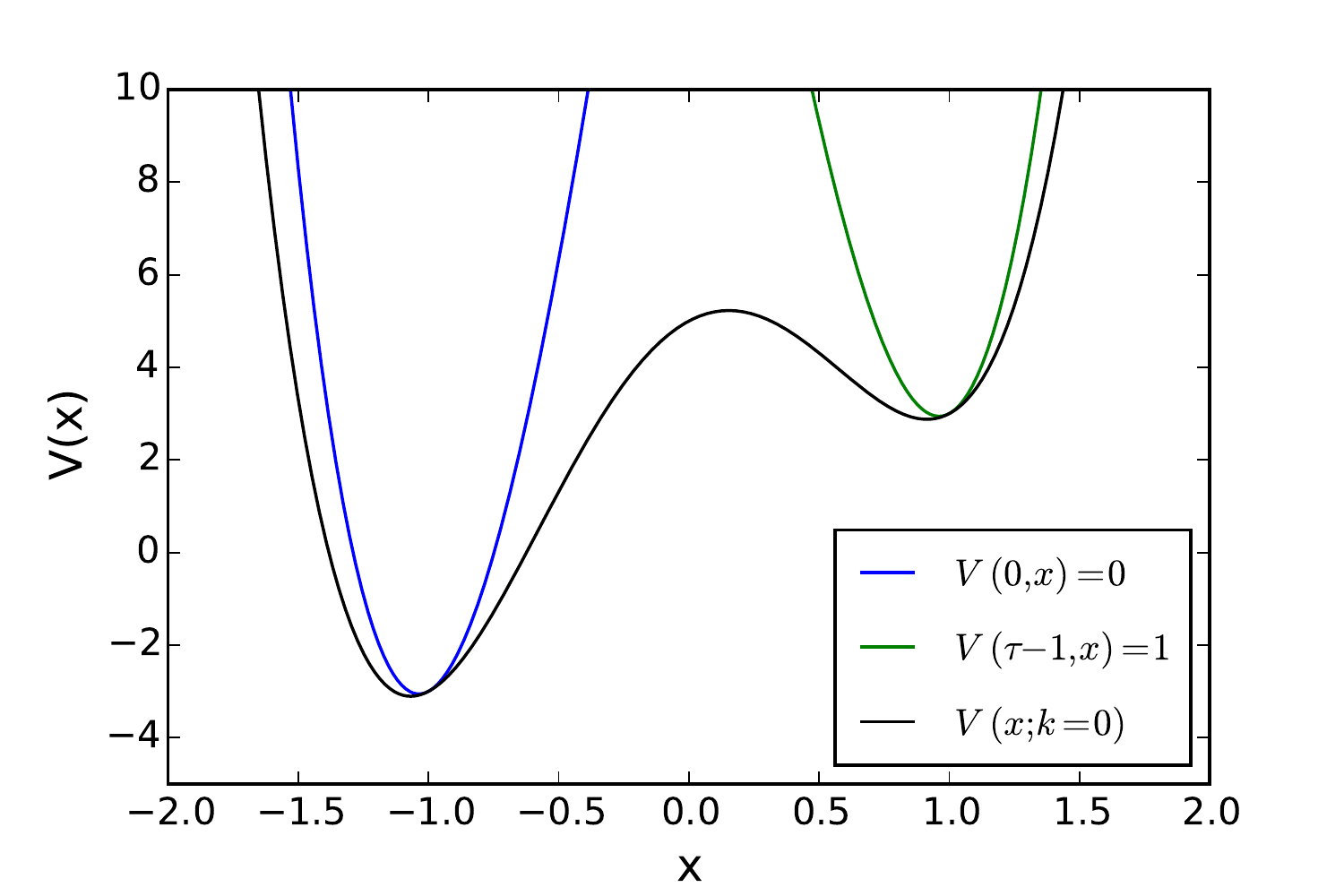}
\caption{$V(0,x)$ (blue) and $V(\tau-1,x)$ (green) for the switching process  used to compute Jarzynski's equality.  For reference, the potential with $k=0$ (black) is also shown.}\label{jar_model}	
\end{figure}

The quantity that we seek to compute is the free energy difference between a particle in a double-well potential that is additionally harmonically restrained with spring constant $k=20$ near $x=-1$  and a particle in the same potential restrained near $x=1$.  The model is adapted from the one presented in \cite{Chipot2007}. Setting $\tau = 501$, for $t<\tau$ we define
 \begin{equation}\label{jarv}
	V(t,x) = 5 \left(x^2 - 1\right)^2 + 3x + k \left(x -  \left(2 t\, dt -1\right)\right)^2
\end{equation} 
where  $dt = 0.001.$
We show $V(0,x)$, $V(\tau-1,x)$, and $V(x;k=0)$ in Figure \ref{jar_model}.  The process $X^{(t)}$ evolves according to \eqref{XforJar}.

The reader may be concerned that the expectation in \eqref{jarzynski_rel} is not immediately of the general form in \eqref{f_ave_dynamical} suitable for an application of NEUS. We apply NEUS as described in Section \ref{sec:gen} to the augmented process  $Z^{(t)} = (X^{(t)}, W^{(t)}).$ To compute the expectation of the left hand side of \eqref{jarzynski_rel} via NEUS, we compute the expectation in \eqref{f_ave_dynamical} with 
\begin{equation} \label{fe_f}
	f(t, Z^{(t)}) = \begin{cases}
		\exp(- W^{(t)}) & \text{if } t = \tau-1 \\
		0 & \text{if } t \neq \tau-1.
	\end{cases}
\end{equation}
The index process $J^{(t)}$ marks transitions between regions of the time $t$ and accumulated work $W^{(t)}$ variables. We discretize the work space in overlapping subsets using the \jt{pyramid form in \eqref{psi_def}}. We use 100 subsets with centers evenly spaced on the interval $[-35.0, 35.0]$ with a width of $\Delta_\phi = 0.6$. We discretize time into 5 discrete nonoverlaping subsets every 100 time steps for a total of 500 subsets. We cap the maximum size of the list representation of  $\{\tilde \gamma_{ij}\}$ at 50 entries using the same scheme as in Section \ref{sec:dipeptide}. 

To assess the accuracy of the NEUS result, we perform 10 independent NEUS simulations. For both NEUS and direct simulations, we prepare an ensemble of 1000 starting states $X^{(0)}$ by performing an unbiased simulation with fixed potential $V(0,x)$ for $10^6$ steps, saving every 1000 steps.  The direct fast-switching simulations start from each of these points and comprise 500 steps of integration forward in time; each trajectory contributes equally to the left hand side of \eqref{jarzynski_rel}.  For the NEUS simulations, the vector $a$ is constructed as in \eqref{a_def}, and trajectories are initialized at $J^{(0)}$ by drawing uniformly from this ensemble. 
We set $K = 100$ and $L=1$, and we perform 500 iterations. Each step in $K$ corresponds to a single step of \eqref{XforJar}. 
As in Section \ref{sec:dipeptide}, we sample only in the restricted distributions where there is at least one point stored in $\tilde \gamma$ from which to restart the dynamics.  

\begin{figure}
\centering
\includegraphics[scale=0.6]{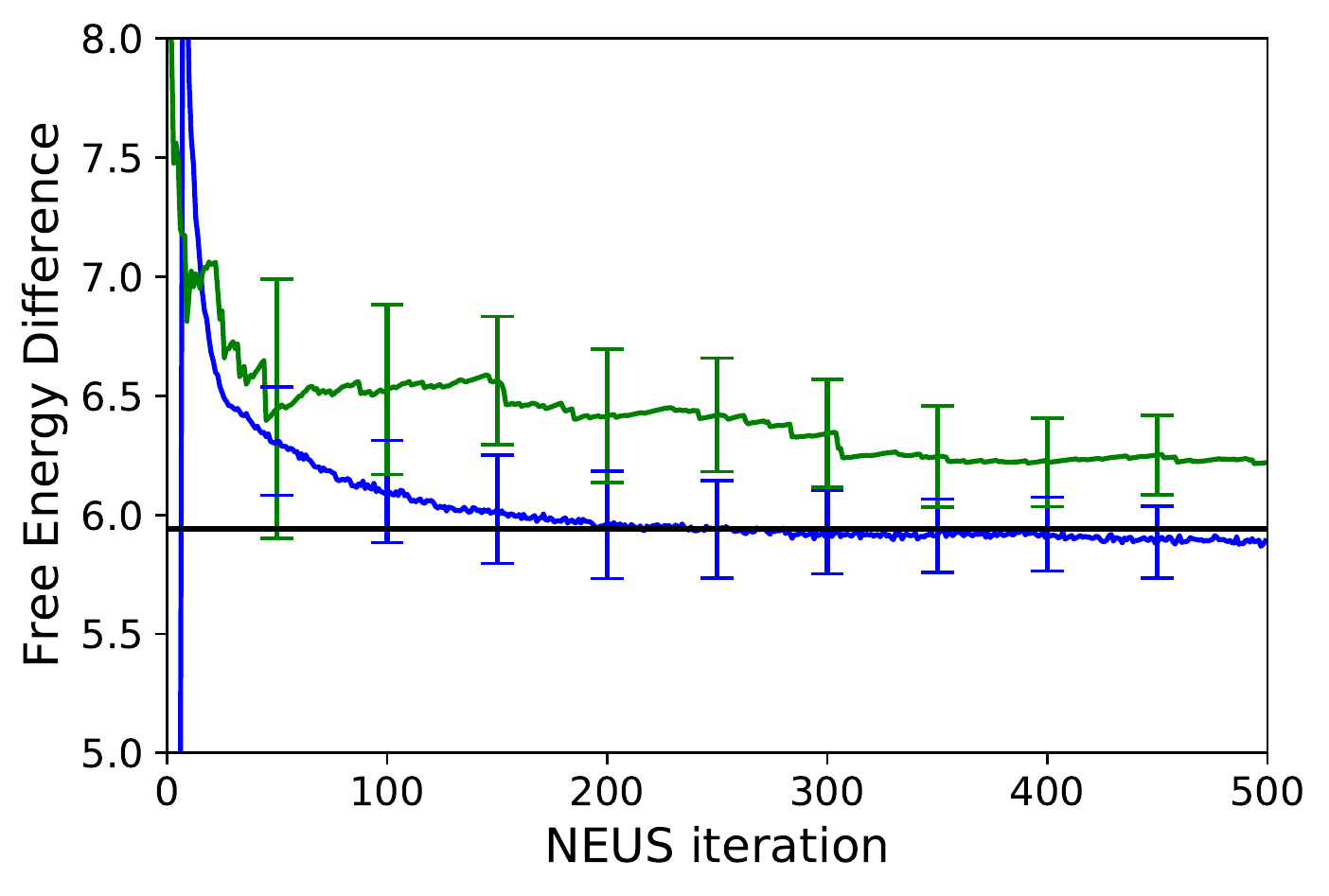}
\caption{Estimate of the free energy computed from NEUS (blue; error bars are computed every 50 iterations and indicate $\pm2.262s/\sqrt{n}$ where $s$ is the standard error estimated from $n=10$ independent NEUS simulations) and from conventional fast-switching simulations (green; error bars are computed every 50 iterations and indicate $\pm2.262s/\sqrt{n}$ where $s$ is the standard error estimated from $n=10$ independent direct simulations). The value computed from numerically integrating the potentials is shown as a black line. For the direct fast-switching simulations, we scale the number of repetitions to the  number of NEUS iterations that are equivalent in computational effort.}\label{jar_estimates}
\end{figure}

\begin{figure}
\centering
\includegraphics[scale=0.61]{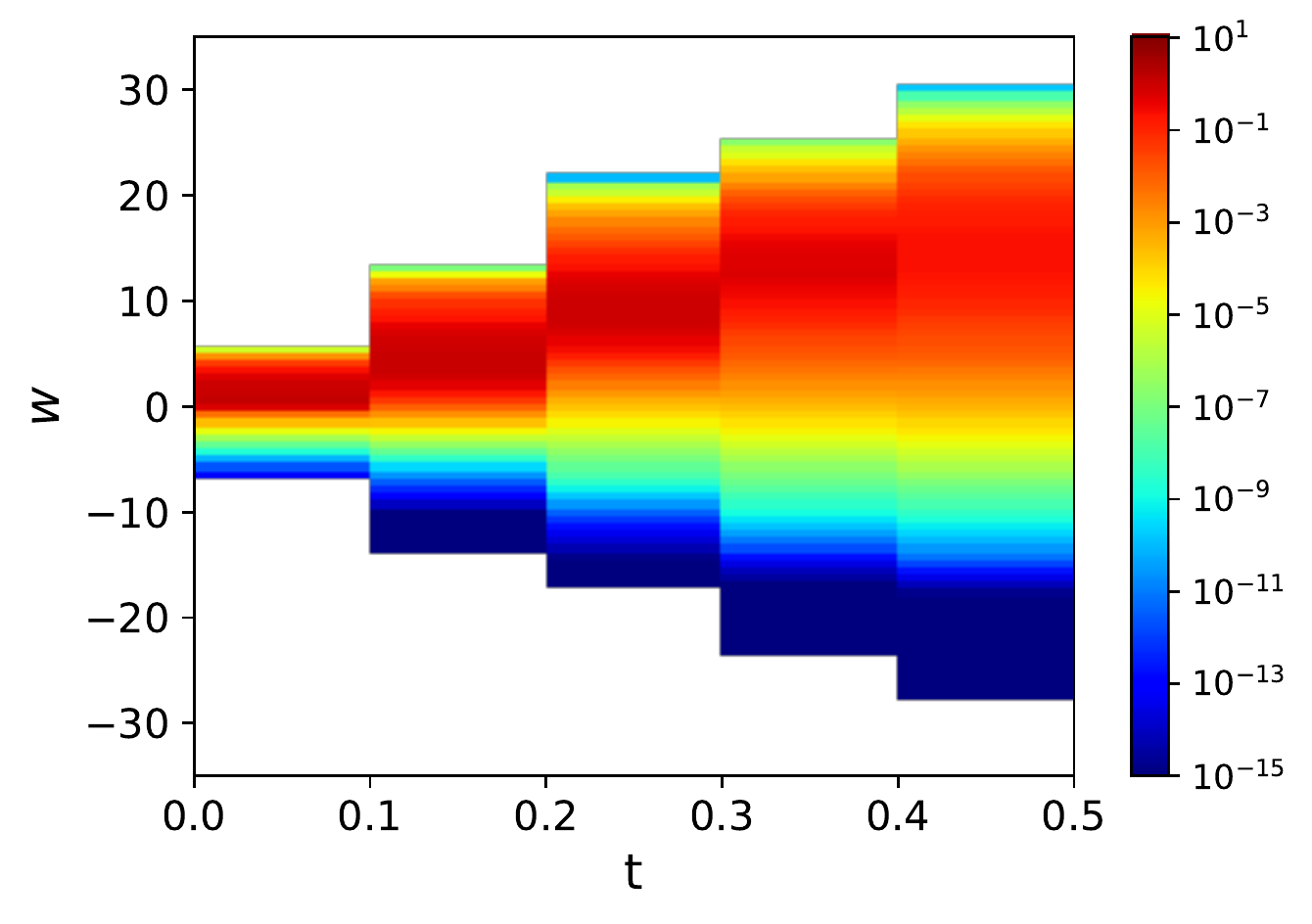}
\includegraphics[scale=0.6]{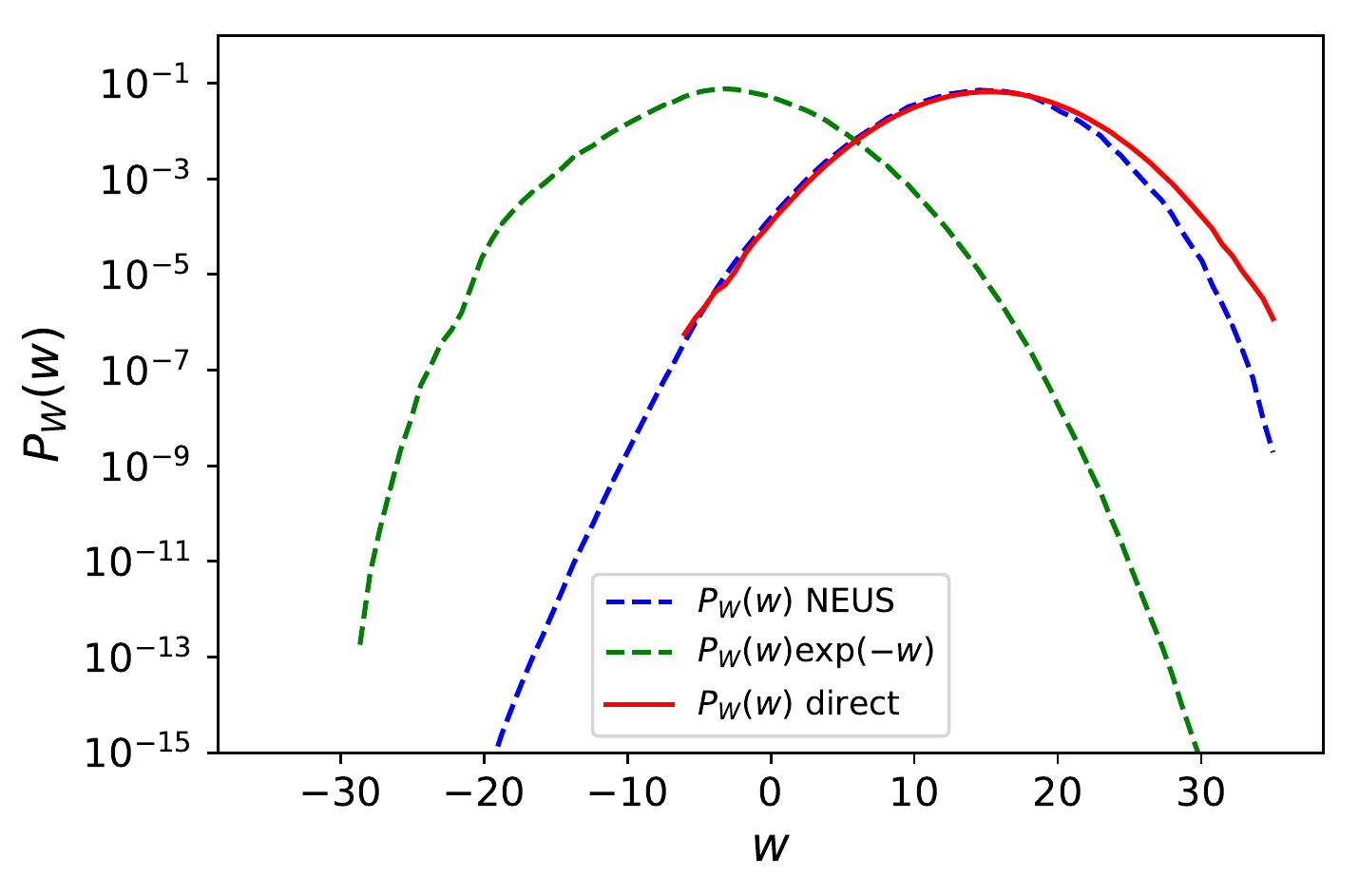}
\caption{Sampling the work with NEUS. (top) The estimate of the dynamic weights, $\bar{z}_{j}$, from the final iteration of the NEUS calculation. White space represents subsets that are not visited in the NEUS calculation. (bottom) The probability density $P_W(w)$ of the accumulated work $W^{(\tau-1)}$ estimated from NEUS (blue dashed line), from direct integration (red solid line) and the exponentially scaled probability density proportional to $P_W(w) \exp(-w)$ estimated from the NEUS calculations (green dashed line). The estimates of $P_{W}(w)$ and $P_W(w) \exp(-w)$ from NEUS (blue dashed line and green dashed line respectively) at each value of $W^{(\tau-1)}$ are computed as an average over the last 10 iterations and then averaged over 10 independent NEUS simulations. The estimate of $P_{W}(w)$ from direct integration (red solid line) is computed as an average over 10 independent direct simulations that are equivalent in computational effort to the 10 independent NEUS simulations. } \label{jar_weights}
\end{figure}

The estimated $\Delta F$  produced from data generated in the last 50 iterations of NEUS is 5.89 (the units are chosen to absorb temperature factors above), which is in excellent agreement with the reference value of 5.94, in contrast to the estimate from direct simulation (Figure \ref{jar_estimates}). 
The left panel of Figure \ref{jar_weights} shows the weights along the time and work axes.  In  the right panel of Figure \ref{jar_weights} we plot  histogram approximations of the density $P_W(w)$ of $W^{(\tau-1)}$ along with the weighted density proportional to $P_W(w) \exp(- w)$. The separation of the peaks of this distribution highlight how NEUS is able to effectively sample the low work tails that contribute significantly to the expectation in the Jarzynski relation in \eqref{jarzynski_rel} but are rarely accessed by the switching procedure in the unbiased simulations. 

\section{Conclusions}
\label{sec:disc}

We describe a trajectory stratification framework for the estimation of  expectations with respect to arbitrary Markov processes. The basis for this framework is the nonequilibrium umbrella sampling method (NEUS) originally  introduced to compute steady state averages. Our development highlights the structural similarities between the nonequilibrium and equilibrium US algorithms and places the NEUS method within the general context of stochastic approximation.   These connections have practical implications for further optimizing the procedure and point the way to a more in depth convergence analysis that will be the subject of future work.

Our development reveals that the basic trajectory stratification approach can be useful well beyond the estimation of stationary averages for time-homogenous Markov processes.  This flexibility is demonstrated in two examples, both involving an expectation over trajectories of finite duration.  In the first example, we show that the probability of first hitting a set within a finite time can be efficiently computed via stratification even when the dynamics start close to a competing absorbing state. In our second example, we use NEUS to stratify a process according to a path-dependent variable, the accumulated work in a nonequilibrium process  appearing in  the Jarzynski equation.  The result is a novel and effective scheme for estimating free energy differences  by enhancing sampling of the tails of the accumulated work distribution.

Our general framework  also suggests  new and exciting applications of trajectory stratification.  For example, with little modification, these methods can be applied to sequential data assimilation applications where the goal is to approximate averages with respect to the conditional distribution  of a hidden signal $X^{(t)}$ given sequentially arriving observations (i.e., with respect to the posterior distribution).  In high-dimensional settings (e.g., weather forecasting) the only practical alternatives are limited to providing information  about only the mode of the posterior distribution (i.e., variational methods) or involve uncontrolled and often unjustified approximations (i.e., Kalman-type schemes).  The approach that we present here opens the door to efficient data assimilation, machine learning, and, more generally, new forms of analysis of complex dynamics.

\begin{appendix}

\section{An Alternative $F$}
\label{Fflux}

Here we present an alternative construction of the stochastic matrix $F$ (Section \ref{sec:eus}) that more closely aligns with the nonequilibrium version of the algorithm presented in Section \ref{sec:gen}.  Suppose that one has available a transition distribution $p(dy\,|\,x)$ for a Markov chain that preserves (or nearly preserves) the target density, $\pi,$ in the sense that
 \begin{equation}
 \pi(dy) = \int_{x \in \mathbb{R}^d} p(dy\,|\,x) \pi(dx).
 \end{equation}
 For example, $p(dy\,|\,x)$ might be the transition density for a number of steps of a Langevin dynamics integrator.
  We can again express the $z_i$ as the solution to an eigenproblem \eqref{eigeq2}
 where now\jt{
  \begin{equation}
 F_{ij} = \int_{y \in \mathbb{R}^d} \int_{x \in \mathbb{R}^d}
\psi_j(y) p(dy\,|\,x)\pi_i(dx).
 \end{equation}}Note that when $\psi_i(x) = \mathbf{1}_{A_i}$ for some partition of space $\{A_i\},$ and $p(dy\,|\,x)$ is reversible with respect to $\pi,$ the entry $F_{ij}$ can be estimated by evolving samples according to $p(dy\,|\,x),$ rejecting any proposed samples that lie outside of $A_i$ (so that $\pi_i$ is preserved), and then counting the number of times the  chain attempts  transitions from set $A_i$ to set $A_j$.  For a closely related approach to approximating certain nonequilibrium quantities see \cite{VandenEijndenVenturoli:2009:MarkovMilestone}.

 \section{Expressions for $\bar G$ and $\bar z$}
 \label{apx:proofs}
 
 In this appendix we establish the identities
 \begin{equation}
  \bar G_{ij} = \begin{cases} \frac{G_{ij}}{1-G_{ii}}, & i\neq j\\
 0, & i=j
 \end{cases}
 \qquad\text{
 and}
 \qquad
 \bar z_j = (1-G_{jj}) z_j
 \end{equation}
 appearing in
 \eqref{barG} and \eqref{barz}.  First, note that the equality $\bar z_i \bar G_{ij} = z_i G_{ij}$ for $i\neq j$ (which follows immediately from the definitions of $\bar z,$ $\bar G,$ $z$, and $G$) together with $ 1-G_{jj} = \bar z_j / z_j$
 implies the expression for $\bar G$ in terms of $G.$   It remains then only to establish the expression for $\bar z$ in terms of $z$ and $G.$  To that end, notice that
\begin{align}
z_j &= \sum_{t=0}^\infty \mathbf{P}\left[J(t)=j,\, t<\tau\right] \nonumber  \\
& = \mathbf{P}\left[J^{(0)}=j\right] +\sum_{t=0}^\infty \mathbf{P}\left[t+1<\tau, J^{(t+1)}=j, J^{(t)}=j\right] \nonumber \\
&\hspace{0.1cm}+  \sum_{t=0}^\infty \mathbf{P}\left[t+1<\tau, J^{(t+1)}=j, J^{(t)}\neq j\right] \nonumber  \\
& = a_j+  z_j G_{jj}\nonumber \\
&\hspace{0.1cm} + \sum_{t=0}^\infty \sum_{\ell=0}^\infty 
\mathbf{P}\left[S^{(\ell+1)}<\tau, S^{(\ell+1)}=t+1, J^{(S^{(\ell+1)})}=j\right] \nonumber  \\
& =  a_j+  z_j G_{jj} + \sum_{\ell=0}^\infty 
\mathbf{P}\left[S^{(\ell+1)}<\tau, J^{(S^{(\ell+1)})}=j\right] \nonumber \\
& = z_j G_{jj}  + \tilde z_j
\end{align}
so that
\begin{equation}
\frac{\tilde z_j}{z_j} = (1-G_{jj}).
\end{equation}

\section{Excursions sample the restricted distributions}
\label{apx:excursions sample restricted dists}

Here, we establish~\eqref{pi_from_pi_bar}.
We have 
\begin{widetext}
\begin{align}
 z_j \pi_j(t,dx) &= {\bf P} [t < \tau, X^{(t)} \in dx, J^{(t)}=j] \nonumber \\
 &= 
 {\bf P} [J^{(0)} = j, t<\sigma(0) \wedge \tau, X^{(t)} \in dx ] + \sum_{s=1}^t {\bf P} [J^{(s)}=j, J^{(s-1)}\neq j, t< \sigma(s) \wedge \tau, X^{(t)} \in dx] \nonumber \\
 &= \sum_{s=0}^t \sum_{\ell=0}^\infty {\bf P} [s=S^{(\ell)}, t <\sigma(s) \wedge \tau, X^{(t)} \in dx ] \nonumber \\
 &= \bar z_j \sum_{s=0}^t \int_y 
 {\bf P}_{s,y,j} 
 \left[ t < \sigma(s) \wedge \tau, X^{(t)} \in dx \right] 
 \bar{\pi}_{j}(s, dy) \nonumber  \\
 &=  
 \bar z_j 
 {\bf P} 
 \left [ t <\rho_j+T_j^{(0)}, Y_j^{(t-T_j^{(0)})} \in dx\right ].
\end{align}
\end{widetext}

\end{appendix}

\section*{Acknowledgments}

The authors would like to thank David Aristoff, James Dama, Jianfeng Lu, Charles Matthews, Erik Thiede, Omiros Papaspiliopoulos, and Eric Vanden-Eijnden for helpful discussions. This research is supported by the National Institutes of Health (NIH) Grant Number
5 R01 GM109455-02. Computational resources were provided by the University of Chicago Research
Computing Center (RCC).

\bibliography{siam-neus}
\setcounter{section}{0}

\clearpage

\onecolumngrid
\begin{center}
{\bf SUPPLEMENTARY MATERIAL}
\end{center}

In this supplementary document we introduce a minimal and analytically tractable Markov model where the terminology and notation of the trajectory stratification framework can be expressed to clearly illustrate the method.   This model does not address the utility of trajectory stratification in practice.
\section{A simple Markov Model}
The model is a discrete Markov process $X^{(t)} \in \{1,2,3,4\}$ with the structure

\vspace{1cm}

{\centering
\hspace{2cm}
\begin{picture}(120,120)(-80,0)
\thicklines
\put(30,20){\framebox(24,24){3}} 
\put(30,100){\framebox(24,24){1}} 
\put(140,20){\framebox(24,24){4}} 
\put(140,100){\framebox(24,24){2}} 

\put(62,114){\vector(1,0){70}} 
\put(62,114){\vector(-1,0){0}} 
\put(90,119){\small 1/2} 

\put(62,30){\vector(1,0){70}} 
\put(62,30){\vector(-1,0){0}} 
\put(90,35){\small 1/2} 

\put(62,37){\vector(1,1){70}} 
\put(62,37){\vector(-1,-1){0}} 
\put(69,78){\small 1/2} 

\put(62,107){\vector(1,-1){70}} 
\put(62,107){\vector(-1,1){0}} 
\put(69,62){\small 1/2} 

\end{picture}}\\
The chain has the transition matrix,
\begin{equation}
	T = \begin{bmatrix}
		0 & \frac{1}{2} & 0 & \frac{1}{2} \\
		\frac{1}{2} & 0 & \frac{1}{2} & 0 \\
		0 & \frac{1}{2} & 0 & \frac{1}{2} \\
		\frac{1}{2} & 0 & \frac{1}{2} & 0 \\
	\end{bmatrix}
\end{equation}
 and initial condition
\begin{equation}
{\bf P} [X^{(0)} = y] = \begin{cases}
	\frac{1}{2} & \text{if} \hspace{0.1cm} y = 1\\
	\frac{1}{2} & \text{if} \hspace{0.1cm} y = 3\\
\end{cases}.
\end{equation}
We consider the case where $\tau=2$. In general, the aim is to compute expectations of the form
\begin{equation}\label{eq:E_dyn}
{\bf E}\left[\sum_{t=0}^{\tau-1}f(t,X^{(t)})\right] = \sum_{i=1}^{n} \bar{z}_i \langle \bar{f} \rangle_{i}
\end{equation}
over a domain, $D$, of time-space pairs. For this model, \[D = (t \in \{0,1\} \times x \in \{1,2,3,4\}).\] 
For now, there is no need to choose a particular form of $f(t,x)$, however the choice of this function is determined from the context of a particular application. 
In this example, we outline how an expectation of the form in \eqref{eq:E_dyn} can be computed by stratification. 

\section{The index process}

We first define the index process $J^{(t)}$ over which we stratify the process $X^{(t)}$. For this model we define the index process 
\begin{equation}
	J^{(t)} =  \begin{cases}
					1  \hspace{0.4cm} \text{if} \hspace{0.1cm} X^{(t)} \in \{1,2\}\\
					2  \hspace{0.4cm} \text{if} \hspace{0.1cm} X^{(t)} \in \{3,4\},
				\end{cases}
\end{equation}
where we have grouped the states as:

\vspace{1cm}

{\centering
\hspace{2cm}
\begin{picture}(120,120)(-80,0)
\thicklines
\put(30,20){\framebox(24,24){3}} 
\put(30,100){\framebox(24,24){1}} 
\put(140,20){\framebox(24,24){4}} 
\put(140,100){\framebox(24,24){2}} 

\put(20,90){\framebox(154,44){}}
\put(184, 110){$J=1$}

\put(20,10){\framebox(154,44){}}
\put(184, 30){$J=2$}

\put(62,114){\vector(1,0){70}}
\put(62,114){\vector(-1,0){0}} 
\put(90,119){\small 1/2} 

\put(62,30){\vector(1,0){70}} 
\put(62,30){\vector(-1,0){0}} 
\put(90,35){\small 1/2} 

\put(62,37){\vector(1,1){70}} 
\put(62,37){\vector(-1,-1){0}} 
\put(69,78){\small 1/2} 

\put(62,107){\vector(1,-1){70}} 
\put(62,107){\vector(-1,1){0}} 
\put(69,62){\small 1/2} 

\end{picture}}\\
For $\tau=2$, the process begins at $X^{(0)} = 1$ or $X^{(0)} = 3$ each with probability one-half and transitions to site 2 or 4 each with probability one-half at time $t=1$. 
At time $t=2$, the process leaves $D$.  There are thus four possible index process sequences; $S^{(0)} = 0$ in all four cases, while the only nonzero realizations of $S^{(\ell)}$ are $S^{(1)} = 1$ for the two cases that contain a switch of index.  

To compute \eqref{eq:E_dyn} by the stratification scheme we outline in Section 3, we first define the flux distributions, $\bar{\pi}_{i}(s,y)$, against which the terms in the left hand side of \eqref{eq:E_dyn} are computed.
The flux distributions for the four-site model are 
\begin{equation}
	\bar{\pi}_i(s,y) = \frac{1}{\bar{z}_i}{\bf P}\left[ S^{(\ell)} = s, s < \tau, X^{(s)} = y, J^{(s)} = i \right] 
\end{equation}
with normalization constants
\begin{align*}
\bar{z}_{1} &= \sum_{\ell=0}^{\infty} {\bf P}[ J^{(S^{(\ell)})} = 1 , S^{(\ell)} < \tau] \\
&= {\bf P}[J^{(S^{(0)})} = 1]  + {\bf P}[J^{(S^{(1)})} = 1, S^{(1)} < \tau] \\
&= \frac{1}{2} + \frac{1}{4} = \frac{3}{4}
\end{align*}
 and similarly, $\bar{z}_2 =\frac{3}{4}$.  We then write the expectation in \eqref{eq:E_dyn} as 
 \begin{equation}
 	 {\bf E} \left[\displaystyle\sum_{t=0}^{\tau-1} f(t,X^{(t)}) \right] = 	\frac{3}{4} \langle \bar{f} \rangle_{1} + \frac{3}{4} \langle \bar{f} \rangle_{2}\\
 \end{equation}
where
\begin{equation}
 \langle \bar{f} \rangle_{i} = \displaystyle\sum_{s=0}^{\infty}  \displaystyle\sum_{y=1}^{4} \displaystyle\sum_{t=0}^{\infty} \displaystyle\sum_{x =1}^{4} f(t,x) {\bf P}[t < \sigma(s) \wedge 2, X^{(t)} = x ] \bar{\pi}_{i}(s,y).
\end{equation}

As in Section 2.2.2 of the main text, the $\bar{z}_{i}$ can be expressed as expectations over the $\bar{\pi}_{i}$. 
Recall that, 
\begin{equation}
	\bar{G}_{ij} = \frac{\sum_{\ell=0}^{\infty} {\bf P} \left[ S^{(\ell + 1)} < \tau, J^{(S^{(\ell+1)})} =j, J^{(S^{(\ell)})} = i \right]}{\bar{z}_{i}}.
\end{equation}
From the definition of $\bar{G}_{ij}$, the exact transition matrix for $\tau=2$ is 
\begin{equation}\label{eq:G_mat}
	\bar{G} = \begin{bmatrix} 
		0 & 1/3 \\
		 1/3 & 0
		\end{bmatrix}
\end{equation} 
and the initial conditions are
\begin{equation}
	a = \begin{bmatrix}
		{\bf P} [ J^{(0)} = 1] \\ 
		{\bf P} [ J^{(0)} = 2] 
	\end{bmatrix} = \begin{bmatrix}
		1/2 \\ 
		1/2
	\end{bmatrix}
\end{equation}
by definition. 
The related affine eigenequation, $\bar{z}^\text{\tiny T} = \bar{z}^\text{\tiny T} \bar{G} + a^\text{\tiny T}$, can be solved to verify that $\bar{z}_1 = \bar{z}_2 = 3/4$. 

So far, we have outlined how \eqref{eq:E_dyn} can be computed from stratification in this four-site model by computing expectations against the two flux distributions, $\bar{\pi}_{1}(s,y)$ and $\bar{\pi}_{2}(s,y)$. 
In the next section, we outline how the NEUS algorithm can be applied to this model to compute \eqref{eq:E_dyn} from the stratification strategy outlined in this section. 

\section{The NEUS fixed-point equations}

In the previous section, we determined the flux distributions, $\bar{\pi}_{i}(s,y)$ and the corresponding weights, $\bar{z}_{i}$, required to compute \eqref{eq:E_dyn} by stratification for the four-site Markov model. 
For this simple model, these terms can be evaluated exactly, and no sampling is required to compute \eqref{eq:E_dyn}. 
In general, however, these terms need to be computed by solving the self-consistent iteration at the heart of the NEUS algorithm. 
In this section we interpret the self-consistent procedure in the context of the four-site Markov model. 

\subsection{The flux distributions}

The exact flux distributions, $\bar\pi_{j}(s,y)$, are related to the conditional flux distributions $\gamma$ by the following identity: 
\begin{equation}
	\bar\pi_{j}(s,y) = \frac{1}{\bar{z}_{j}} \begin{cases}
		\sum_{i\neq j} \bar{z}_{i} \bar{G}_{ij} \gamma_{ij}(s,y) &\text{if} \hspace{0.1cm} s > 0\\
		a_{j} {\bf P} [ X^{(0)} \in y \hspace{0.1cm}|\hspace{0.1cm} J^{(0)} = j ]  & \text{if} \hspace{0.1cm} s = 0
	\end{cases}
\end{equation}
where $\gamma_{ij}$ are the conditional flux distributions for transitions from $J=i$ to $J=j$. 
In general, the conditional flux distributions are 
\begin{equation}
	\gamma_{ij} (s, y) = \frac{1}{\bar{z}_i \bar{G}_{ij}} \sum_{\ell = 0}^{\infty} {\bf P} [s = S^{(\ell +1)} < \tau, J^{(S^{(\ell)})} = i, J^{(s)} = j, X^{(s)} = y]
\end{equation}
or, for this simple model,
\begin{equation}
	 \gamma_{21}(s,y) = \begin{cases}
 			1 & \text{if} \hspace{0.1cm} (s,y) = (1,2)\\
 			0 & \text{otherwise}
 \end{cases}
\end{equation} 
and 
\begin{equation}
	 \gamma_{12}(s,y) = \begin{cases}
 			1 & \text{if} \hspace{0.1cm} (s,y) = (1,4)\\
 			0 & \text{otherwise}.
 \end{cases}
\end{equation}
 Therefore, noting that 
\begin{equation}
	\frac{\bar{z}_2 \bar{G}_{21}}{\bar{z}_{1}}  = \frac{1}{3}
\end{equation}
and 
\begin{equation}
	\frac{a_{1}}{\bar{z}_{1}}  = \frac{2}{3},
	\label{p}
\end{equation} 
the exact flux distributions are
\begin{equation}\label{eq:pi_bar1}
	\bar{\pi}_{1}(s,y) = \begin{cases}
		\frac{1}{3} & \text{if} \hspace{0.1cm} (s,y) = (1,2) \\
		\frac{2}{3} & \text{if} \hspace{0.1cm} (s,y) = (0,1) \\
		0 & \text{otherwise}
	\end{cases}
\end{equation}
and
\begin{equation}
	\bar{\pi}_{2}(s,y) = \begin{cases}
		\frac{1}{3} & \text{if} \hspace{0.1cm} (s,y) = (1,4) \\
		\frac{2}{3} & \text{if} \hspace{0.1cm} (s,y) = (0,3) \\
		0 & \text{otherwise}.
	\end{cases}
\end{equation}

As discussed in Section 3, we interpret NEUS as a stochastic approximation algorithm to solve the  deterministic fixed point equation 
\begin{equation}\label{eq:fixed-point}
	(\mathcal{G}(\bar{G}, \gamma), \Gamma(\bar{G}, \gamma)  )= (\bar{G}, \gamma),
\end{equation}
where the matrix $\bar{G}$ and conditional flux distributions $\gamma$ are the fixed points of a pair of maps, $\mathcal{G}$ and $\Gamma$, that take $\tilde G$ and $\tilde \gamma$ as arguments and return a new approximation to $\bar{G}$ and $\gamma$ respectively. Here, however,  we interpret the deterministic fixed point iteration
\[
(\mathcal{G}(\tilde G(m), \tilde \gamma(m)), \Gamma(\tilde G(m), \tilde \gamma(m))  )= (\tilde G(m+1), \tilde \gamma(m+1)),
\]
 for the four-site model and show that, for this model, this iteration can be expressed in a single variable representing the relative contribution of the fluxes $\tilde z_{i}(m) \tilde G_{ij}(m)$ where $\tilde z(m)$ solves $\tilde z^\text{\tiny T}(m) = \tilde z^\text{\tiny T}(m) \tilde G(m) + a^\text{\tiny T}$. 

We define the Markov process $\mathcal{Y}^{(r)}_{j}(\tilde{G}(m), \tilde{\gamma}(m))$ that samples each approximate restricted distribution $\Pi_{j}(t,x;\tilde{G}(m), \tilde{\gamma}(m))$. The process $\mathcal{Y}^{(r)}_{j}(\tilde{G}(m), \tilde{\gamma}(m))$ generates a set of Markov processes called excursions where each excursion is a finite segment of a trajectory of the underlying process $(t,X^{(t)}, J^{(t)})$ with $J = j$. 
Each excursion evolves with the underlying process $(X^{(t)}, J^{(t)})$ forward in time until the process transitions from $J = j$ to $J=i$ or hits $t = \tau$, at which point a new initial point is drawn from $\bar{\Pi}_{j}(t,x;\tilde{G}(m), \tilde{\gamma}(m))$ and the process is continued from the new initial point. 
For clarity, we use the notation $\mathcal{Y}^{(r)}_{j}(m)$ to denote the chain $\mathcal{Y}^{(r)}_{j} (\tilde{G}(m), \tilde{\gamma}(m))$. 

For the four-site model, the conditional flux distributions are determined exactly by sampling $\mathcal{Y}_{j}^{(r)}(m)$, i.e., $\tilde{\gamma}_{21} (1,2) = \gamma_{21} (1,2) =1$ and $\tilde{\gamma}_{12} (1,4) = \gamma_{21} (1,4) = 1$.
Therefore, the function $\bar{\Pi}_{j}(t,x;\tilde{G}(m), \tilde{\gamma}(m))$ depends only on determining the relative fluxes, $\tilde z_{i}(m) \tilde G_{ij}(m)$. 
Specifically let 
\begin{equation}\label{eq:p_def}
p(m) = \frac{a_{1}} { \tilde z_{1}(m)}
 \end{equation}
be the relative weight of the $s=0$ time contribution to $\bar{\Pi}_{1}(t,x;\tilde{G}(m), \tilde{\gamma}(m))$ at the $m$th iteration.   From \eqref{p} we see that  $p(m)$ should converge to $2/3$ as $m$ increases.
Because the underlying dynamics are symmetric for transitions between the index process $J$, we derive this fixed point equation only for the single number $p(m)$ described above for $J=1$ since the analogous ratio for $J=2$ is equivalent at each iteration. 
In the following we express \eqref{eq:fixed-point} as a fixed point expression in $p(m)$. 

To compute the expectation of $\tilde{G}_{12}(m)$ from $\mathcal{Y}_{1}^{(r)}(m)$, we enumerate all possible excursions of $\mathcal{Y}^{(r)}_{1}(m)$. We assume here that at each iteration $m$ we compute expectations from the process $\mathcal{Y}_{1}^{(r)}(m)$ exactly. For brevity, we momentarily drop the iteration index $m$ from our notation. The three possible excursions for $\mathcal{Y}^{(r)}_{1}$ are as follows:
\begin{itemize}
	\item $\mathcal{Y}^{(r)}_1$ draws initial time and position pair $(s,y) = (0, 1)$ from $\bar{\Pi}_{1}$ with probability $p$ and transitions from $1 \to 4$ with probability $1/2$. This excursion has overall probability $p/2$. 
	\item $\mathcal{Y}^{(r)}_1$ draws initial time and position pair $(s,y) = (0, 1)$ from $\bar{\Pi}_{1}$ with probability $p$ and transitions from $1 \to 2$ with probability $1/2$. This excursion has overall probability $p/2$. 
	\item $\mathcal{Y}^{(r)}_1$ draws initial time and position pair $(s,y) = (1, 2)$ from $\bar{\Pi}_{1}$ with probability $1-p$.
\end{itemize}

Only the first of the excursions above contributes to the expectation of $\tilde{G}_{12} (m)$ and occurs with probability $p(m)/2$. Therefore $\tilde{G}_{12}(m) = p(m) / 2$. By symmetry, $\tilde{G}_{21} (m)$ has the same form. Therefore the matrix 
\begin{equation}
	\tilde{G}(m+1) = \begin{bmatrix}
		0 & p{(m)} / 2\\
		p{(m)} / 2 & 0\\
	\end{bmatrix}. 
\end{equation}
The weights $\tilde z(m+1)$ are computed by solving $\tilde z(m+1)^{T} = \tilde z(m+1)^{T} \tilde G(m+1)^{T} + a^{T}$ which gives the solution $\tilde z_{1}(m+1) = \tilde z_{2}(m+1) = 1/(2-p(m))$. 

Finally, we substitute this solution into  \eqref{eq:p_def} to express  $p(m+1)$ in terms of $p(m)$:
\begin{equation}\label{eq:p_fixed}
	p(m+1) = 1-\frac{p(m)}{2}.
\end{equation}
Iteration of this relation converges to the value 2/3, as desired.
 
For the four-site model, the NEUS fixed point equation can be reduced to a fixed point equation in a single variable that represents the relative weight of the contributions to $\bar\pi_{i}$. In general, the terms in the fixed point expression cannot be computed exactly and a stochastic approximation strategy like NEUS is required. 

\end{document}